\documentclass[twocolumn]{aastex62}

\usepackage{newtxmath}

\shorttitle{Parametric SFH Models}
\shortauthors{Carnall et al.}

\begin{document}

\title{How to measure galaxy star-formation histories I: Parametric models}

\correspondingauthor{Adam C. Carnall}
\email{adamc@roe.ac.uk}

\author{Adam C. Carnall}
\affiliation{Institute for Astronomy, University of Edinburgh, Royal Observatory, Edinburgh EH9 3HJ, UK}
\affiliation{SUPA, Scottish Universities Physics Alliance}

\author{Joel Leja}
\affiliation{NSF Astronomy and Astrophysics Postdoctoral Fellow}
\affiliation{Harvard-Smithsonian Center for Astrophysics, 60 Garden St. Cambridge, MA 02138, USA}

\author{Benjamin D. Johnson}
\affiliation{Harvard-Smithsonian Center for Astrophysics, 60 Garden St. Cambridge, MA 02138, USA}

\author{Ross J. McLure}
\affil{Institute for Astronomy, University of Edinburgh, Royal Observatory, Edinburgh EH9 3HJ, UK}
\affiliation{SUPA, Scottish Universities Physics Alliance}

\author{James S. Dunlop}
\affil{Institute for Astronomy, University of Edinburgh, Royal Observatory, Edinburgh EH9 3HJ, UK}
\affiliation{SUPA, Scottish Universities Physics Alliance}

\author{Charlie Conroy}
\affiliation{Harvard-Smithsonian Center for Astrophysics, 60 Garden St. Cambridge, MA 02138, USA}

\begin{abstract}

\noindent Parametric models for galaxy star-formation histories (SFHs) are widely used, though they are known to impose strong priors on physical parameters. This has consequences for measurements of the galaxy stellar-mass function (GSMF), star-formation-rate density (SFRD) and star-forming main sequence (SFMS). We investigate the effects of the exponentially declining, delayed exponentially declining, lognormal and double power law SFH models using \textsc{Bagpipes}. We demonstrate that each of these models imposes strong priors on specific star-formation rates (sSFRs), potentially biasing the SFMS, and also imposes a strong prior preference for young stellar populations. We show that stellar mass, SFR and mass-weighted age inferences from high-quality mock photometry vary with the choice of SFH model by at least 0.1, 0.3 and 0.2 dex respectively. However the biases with respect to the true values depend more on the true SFH shape than the choice of model. We also demonstrate that photometric data cannot discriminate between SFH models, meaning it is important to perform independent tests to find well-motivated priors. We finally fit a low-redshift, volume-complete sample of galaxies from the Galaxy and Mass Assembly (GAMA) Survey with each model. We demonstrate that our stellar masses and SFRs at redshift, $z\sim0.05$ are consistent with other analyses. However, our inferred cosmic SFRDs peak at $z\sim0.4$, approximately 6 Gyr later than direct observations suggest, meaning our mass-weighted ages are significantly underestimated. This makes the use of parametric SFH models for understanding mass assembly in galaxies challenging. In a companion paper we consider non-parametric SFH models.

\end{abstract}


\section{Introduction} \label{sect:intro}

One of the most important processes driving the evolution of galaxies is star formation. This means that the stellar masses and star-formation rates (SFRs) of galaxies are two of their most fundamental properties. Measurements of these quantities underpin many of the most important results in the study of galaxy formation, such as the redshift evolution of the galaxy stellar-mass  function (GSMF; e.g. \citealt{Tomczak2014}), cosmic star-formation-rate density (SFRD; e.g. \citealt{Madau2014}) and the galaxy star-forming main sequence (SFMS; e.g. \citealt{Speagle2014}).

To measure these quantities we rely on modelling and fitting the observed spectral energy distributions (SEDs) of galaxies (see \citealt{Conroy2013}), using methods ranging from monochromatic SFR indicators and single-colour mass-to-light relationships to full spectral fitting. 

Models used to fit galaxy SEDs normally include a star-formation history (SFH). The fitted SFH is then used to derive the SFR and stellar mass (as distinct from the total stellar mass formed; see Section \ref{sect:priors}), as well as other quantities of interest such as the specific SFR (sSFR) and mass-weighted age. This means that the priors on these quantities are not set explicitly, but instead are set implicitly by the priors applied to the SFH. Because of this, SFH priors affect results obtained for many of the most fundamental galaxy properties.

As the choice of prior is subjective, it is generally desirable for the conclusions reached to be as insensitive to the prior assumptions made as possible. For the prior to be of minimal importance, two conditions must be met. Firstly, the model being fitted must be capable of accurately describing the data-generating processes. Secondly, the data must be strongly constraining on the values of all model parameters. The problem of fitting SFHs to galaxy SEDs is particularly challenging because, in general, neither of these conditions are met.

In this case the data-generating processes encompass the entire physics of galaxy formation and evolution, and are thus extremely complex. It has been shown through simulations of galaxy formation (e.g. \citealt{Dave2016}; \citealt{Nelson2018}) that these processes can give rise to a huge diversity of complex galaxy SFHs, presenting a significant modelling challenge. 

Progress has been made in addressing this challenge, however there is still much debate in the literature as to the best approach. Three different families of models are in common usage, all of which impose substantially different priors on SFHs, both in terms of the range of allowed shapes and the relative prior weights assigned to different allowed shapes. Firstly, parameterised SFH models, which are the subject of this work (see Section \ref{sect:models}), secondly non-parametric models (e.g. \citealt{CidFernandes2005}; \citealt{Ocvirk2006}; \citealt{Tojeiro2007}; \citealt{Cappellari2017}; \citealt{Leja2017}; \citealt{Chauke2018}), and thirdly models drawn directly from simulations (e.g. \citealt{Brammer2008}; \citealt{Pacifici2012}). At the core of this debate is a trade-off between flexibility and computational tractability, with more flexible models generally being more computationally intensive to fit.

Even if a SFH model can be defined which is both computationally tractable and flexible enough to encompass the inherent complexity of galaxy SFHs, the  priors assumed for its parameters will still be important unless all of the model parameters are well constrained by the data. However, as was extensively demonstrated by \cite{Ocvirk2006}, the problem of inferring SFHs from galaxy SEDs is poorly conditioned, meaning even small perturbations of the data can lead to large perturbations of the inferred SFH. The main underlying reason for this is the strong evolution of the mass-to-light ratios of stellar populations with age, meaning that the early-time evolution of the SFH has little effect on the observed SED at later times if star-formation is ongoing.

The situation therefore is one in which inferences made about key galaxy physical parameters are highly sensitive to the SFH prior (e.g. \citealt{Wuyts2009, Wuyts2011}; \citealt{Lee2009}; \citealt{McLure2011}; \citealt{Pforr2012}; \citealt{Mobasher2015}; \citealt{Salmon2015}; \citealt{Iyer2017}; \citealt{Carnall2018}). SFH priors are therefore a subject of prime importance for all SED fitting analyses.

In this context it is important to critically evaluate SFH priors to see whether the priors they impose on parameters of interest are well-motivated. Two significant risks are the over-interpretation of data if observational uncertainties are not carefully propagated, and the unintentional imposition of informative priors, which can lead to strong posterior constraints even in the absence of strongly constraining data. It is possible that issues of this nature contribute to the known tensions between the observed GSMF, SFRD and SFMS (e.g. \citealt{Madau2014}; \citealt{Leja2015}; \citealt{Pacifici2015}).

Recently, significant advances in statistical and computational techniques (e.g. \citealt{Skilling2006}; \citealt{Feroz2009, Feroz2013}; \citealt{Goodman2010}; \citealt{Acquaviva2011}; \citealt{Foreman-Mackey2013}) have made it possible to rapidly fit complex galaxy SED models to data within a fully Bayesian framework, including full control of the applied priors. A new generation of galaxy SED fitting tools has been built to exploit this, such as \textsc{Beagle} \citep{Chevallard2016}, \textsc{Prospector} (\citealt{Leja2017}; Johnson et al. in preparation) and \textsc{Bagpipes} \citep{Carnall2018}. These codes allow detailed analyses of the effects of SFH priors, and direct comparisons between results obtained under different assumptions.

In this work, we use Bayesian Analysis of Galaxies for Physical Inference and Parameter EStimation (\textsc{Bagpipes}) to conduct an investigation into the priors imposed by four commonly used parametric SFH models: exponentially declining, delayed exponentially declining, lognormal and double power law. In a companion paper \citep{Leja2018a} we perform similar analyses for non-parametric SFH models.

In Section \ref{sect:models} we introduce the parametric SFH models. In Section \ref{sect:priors} we consider the effects of these models in the limit of weakly constraining data by discussing the priors they impose on physical parameters, in particular sSFR and mass-weighted age/formation time. In Section \ref{sect:mocks}, we consider the biases introduced in the case of fitting these models to high signal-to-noise ratio (SNR) broad-band photometric data. This is achieved by constructing and fitting a simple mock galaxy catalogue, designed to span a wide range of scenarios for the formation of galaxies. We also discuss how well these mock data can discriminate between different SFH models.

Finally, in Section \ref{sect:madau}, we perform a long-discussed consistency check (e.g. \citealt{Heavens2004}) by using these four models to fit a volume-complete sample of local galaxies and comparing the redshift evolution of the cosmic SFRD inferred from their SFHs to the relationship obtained by measuring SFRs across cosmic time by \cite{Madau2014}. For this analysis we use a sample in the redshift interval $0.05 < z < 0.08$ with high-quality photometric data and spectroscopic redshifts from the Galaxy and Mass Assembly (GAMA) Survey \citep{Driver2009, Driver2016, Baldry2018}. 

\begin{table*}
  \caption{Parameters and prior distributions for each SFH model. Logarithmic priors are uniform in log$_{10}$ of the parameter.}
  \begin{center}
\begingroup
\setlength{\tabcolsep}{10pt} 
\renewcommand{\arraystretch}{1.2} 
\begin{tabular}{lllll}
\hline
Model & Parameter & Symbol / Unit & Range & Prior\\
\hline
Exponentially declining / & Start time & $T_0$ / Gyr  & (0, $t_\mathrm{obs} - 0.1$) & uniform \\
Delayed exponentially declining & Timescale & $\tau$ / Gyr  & (0.3, 10) & uniform\\
 \hline
Lognormal & Peak time & $t_\mathrm{max}$ / Gyr  & (0.1, 15) & uniform \\
 & FWHM & $T_\mathrm{FWHM}$ / Gyr  & (0.1, 20) & uniform \\
 \hline
Double power law & Falling slope & $\alpha$ & (0.1, 1000) & logarithmic\\
 & Rising slope & $\beta$ & (0.1, 1000) & logarithmic\\
 & Turnover & $\tau$ / Gyr & (0.1, $t_\mathrm{obs}$) & uniform \\
 \hline
Global & Normalisation & $M_\mathrm{formed}$ / $\mathrm{M_\odot}$ & (1, $10^{13}$) & logarithmic\\
 \hline
\end{tabular}
\endgroup
\end{center}
\label{table:priors}
\end{table*}

We assume $\Omega_M = 0.3$, $\Omega_\Lambda = 0.7$ and $H_0$ = 70 $\mathrm{km\ s^{-1}\ Mpc^{-1}}$. All times, $t$, are measured forwards from the beginning of the Universe such that $t(z)$ is the age of the Universe at redshift $z$. We assume a \cite{Kroupa2002} initial mass function (IMF).

\section{Parametric SFH models}\label{sect:models}

Parametric models approximate galaxy SFHs using simple functional forms, typically involving $2-3$ shape parameters plus a normalisation. These are, in some respects, the least flexible option for fitting SFHs, imposing strong prior limitations on the range of allowable shapes. However their relative speed and simplicity of fitting means that they are widely used. It has also been demonstrated that complex SFHs from simulations can be reasonably well described by parametric models (e.g. \citealt{Simha2014}, \citealt{Diemer2017}).

The four widely used parametric SFH models we consider (and some recent examples of their use in the literature) are: exponentially declining \citep{Mortlock2017, Wu2018, McLure2018}, delayed exponentially declining \citep{Ciesla2017, Chevallard2017}, lognormal \citep{Diemer2017, Cohn2018} and double power law \citep{Ciesla2017, Carnall2018}. For brevity we will refer to them as tau, delayed, lognormal and DPL models respectively.

The models are introduced in the following four subsections; the prior probability densities we assume for their parameters are reported in Table \ref{table:priors}. We have chosen a simple set of prior probability densities in each case to act as a basis for comparison. The effects of changing these will be discussed in Section \ref{subsect:priors_ppds}. In all cases we normalise the SFH by the total stellar mass formed, $M_\mathrm{formed}$ at the time of observation, $t_\mathrm{obs}$. We assign a logarithmic prior to $M_\mathrm{formed}$, which is the minimally informative prior when the uncertainty spans several orders of magnitude (e.g. \citealt{Simpson2017}).

\subsection{Exponentially declining}

Exponentially declining SFHs (tau models) are probably the most commonly applied SFH model. They assume that star-formation jumps from zero to its maximum value at some time $T_0$, after which star-formation declines exponentially with some timescale $\tau$,

\begin{equation}\label{eqn:tau}
\quad \mathrm{SFR}(t)\ \propto\ \begin{cases}
    \ \ \ \mathrm{exp}\bigg(-\dfrac{t - T_0}{\tau}\bigg) & \qquad  t > T_0\\
    \ \ \ 0 & \qquad  t < T_0.
\end{cases}
\end{equation}

\noindent Tau models are often used as a fiducial model against which others can be compared, however they have been shown to become less appropriate at higher redshifts (e.g. \citealt{Reddy2012}) as they cannot reproduce rising SFHs. They have also been shown to produce biased estimates of stellar mass, SFR and mass-weighted age when used to fit mock observations of simulated galaxies (e.g. \citealt{Simha2014}; \citealt{Pacifici2015}; \citealt{Carnall2018}). The priors listed in Table \ref{table:priors} for this model are adapted from those used by \cite{Wuyts2011}.

\subsection{Delayed exponentially declining}

A simple extension of exponentially declining SFHs are delayed exponentially declining SFHs (delayed models). Multiplying the tau-model SFR by the time since $T_0$ removes both the discontinuity in SFR at $T_0$ and the condition that star formation can only decline after that point. This results in a more flexible and more physical model, which can reproduce rising SFHs if $\tau$ is large. The SFR is now described by

\begin{equation}\label{eqn:del}
\quad \mathrm{SFR}(t)\ \propto\ \begin{cases}
    \ \ \  (t - T_0)\ \mathrm{exp}\bigg(-\dfrac{t - T_0}{\tau}\bigg) & \quad  t > T_0\\
    \ \ \ 0 & \quad  t < T_0.
\end{cases}
\end{equation}

\noindent For our delayed model we apply the same prior probability densities for $\tau$ and $T_0$ as for our tau model.

\subsection{Lognormal}

Lognormal models  for individual galaxy SFHs were postulated by \cite{Gladders2013}, based, in part, on the evolution of the cosmic SFRD being well fitted by the lognormal function. The SFR is described by

\begin{equation}\label{eqn:lognorm}
SFR(t)\ \propto\ \frac{1}{t}\ \mathrm{exp} \bigg( -\frac{\big(\mathrm{ln}(t) - T_0\big)^2}{2\tau^2} \bigg)
\end{equation}

\noindent where $\tau$ and $T_0$ are free parameters. Because these parameters do not have intuitive interpretations (e.g. star-formation does not peak at $ t = e^{T_0}$) we follow  \cite{Diemer2017} in reparameterising in terms of $t_\mathrm{max}$, the time at which star-formation peaks, and $T_\mathrm{FWHM}$, the full-width at half maximum of the SFH. We constrain $t_\mathrm{max}$ to be less than 15 Gyr after the beginning of the Universe and $T_\mathrm{FWHM}$ to be less than 20 Gyr in order to limit the prior volume containing models with very large $t_\mathrm{max}$ and $T_\mathrm{FWHM}$ which have almost identical shapes at times earlier than the $z=0$ age of the Universe.

\subsection{Double power law}

The double-power-law (DPL) function introduces another free parameter in order to separate the rising and declining phases of the SFH, which are modelled by two separate power-law slopes. This function has been shown to provide a good description of the redshift evolution of the cosmic SFRD \citep{Behroozi2013, Gladders2013}, as well as producing good fits to SFHs from simulations (e.g. \citealt{Pacifici2016, Diemer2017, Carnall2018}). The functional form is

\begin{equation}\label{eqn:DPL}
\mathrm{SFR}(t)\ \propto\ \Bigg[\bigg(\frac{t}{\tau}\bigg)^{\alpha} + \bigg(\frac{t}{\tau}\bigg)^{-\beta}\Bigg]^{-1}
\end{equation}

\noindent where $\alpha$ is the falling slope, $\beta$ is the rising slope and $\tau$ is related to (but not the same as) the peak time. The priors reported in Table \ref{table:priors} are based on those used by \cite{Carnall2018}.

\begin{figure*}[ht]
	\includegraphics[width=\textwidth]{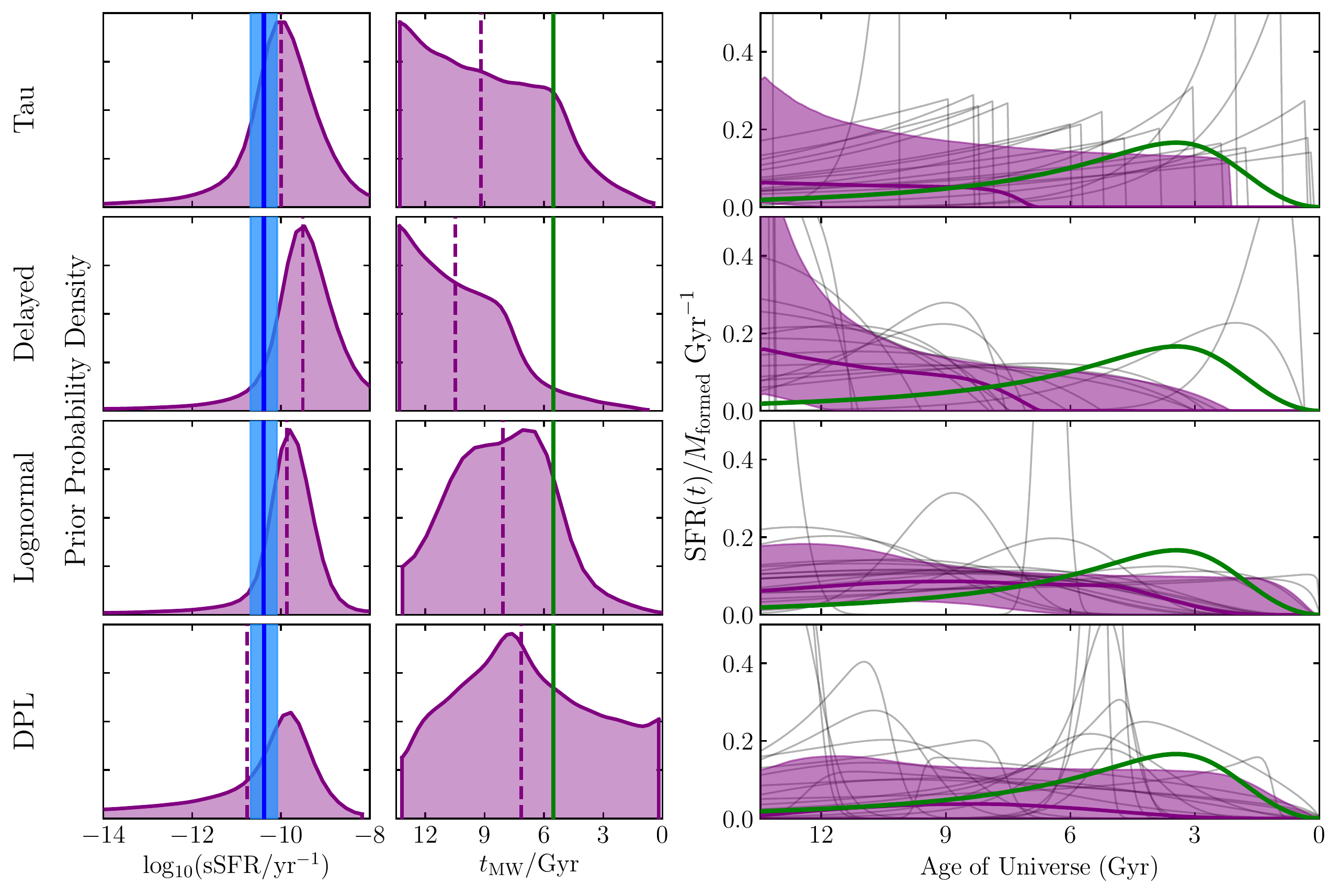}
    \caption{Prior probability densities for sSFR, mass-weighted formation time and SFH shape imposed by each of our four parametric SFH models (see Section \ref{sect:models} and Table \ref{table:priors}). To the left, the prior medians are shown as dashed purple lines. The $z=0$ SFMS of \protect \cite{Speagle2014} for a representative stellar mass of $M_* = 10^{10.5}\mathrm{M_\odot}$ (close to the characteristic mass of the GSMF) is shown on the far-left panels as a blue solid line. The blue shaded region shows a scatter of 0.3 dex. The mass-weighted formation time for the stars in the Universe, derived from the \cite{Madau2014} SFRD curve, is shown as a solid green line on the centre-left panels. To the right, the solid purple line is the prior median and the shaded region shows the 16\textsuperscript{th}--84\textsuperscript{th} percentiles. A sample of draws from each prior is shown in grey. The \cite{Madau2014} SFRD curve is shown in green.}\label{fig:fiducial_priors}
\end{figure*}

\section{Priors on physical parameters}\label{sect:priors}

As described in Section \ref{sect:intro}, in most galaxy SED fitting analyses, the priors on the parameters of interest (e.g. stellar mass, star-formation rate, mass-weighted age) are not set explicitly, instead being set implicitly by the priors applied to the SFH. In this section we use \textsc{Bagpipes} to sample the prior distributions listed in Table \ref{table:priors} for the models described in Section \ref{sect:models} and report and discuss the priors imposed on parameters of interest.

For each draw from the prior, we obtain the stellar mass, $M_*$ (which is the total mass in stars and remnants at the time of observation), by integrating the SFH multiplied by the mass-return fraction as a function of stellar age. This is distinct from the normalisation of the SFH, $M_\mathrm{formed}$ described in Section \ref{sect:models}, which is the total stellar mass formed. We also average over the most recent 100 Myr of the SFH to obtain an estimate of the current SFR, which we denote SFR$_{100}$.

The dependence of $M_*$ on the shape of the SFH is relatively weak, meaning that the prior on $M_*$ is largely independent of the parametric form used. The $M_*$ prior instead closely mirrors the prior placed on $M_\mathrm{formed}$, typically with an offset of $\sim 0.2$ dex (the return fraction). We therefore do not discuss stellar masses in this section, however it should be noted that the SFH prior can still bias measurements of $M_*$ if the model cannot reproduce the true SFH shape (see Section \ref{sect:mocks}).

The SFR prior has significant dependencies on both the SFH model and the stellar mass. In order to isolate the effect of the SFH model, we will consider the prior on the specific star-formation rate, sSFR, which we calculate by sSFR = SFR$_{100}$/$M_*$. 

The second physical parameter we consider in this section is the mass-weighted formation time, $t_\mathrm{MW}$. This corresponds to the more commonly used mass-weighted age, but is measured forwards from the beginning of the Universe to maintain consistency between objects at different observed redshifts. We calculate this by

\begin{equation}
t_\mathrm{MW} = \frac{\int_{0}^{t_\mathrm{obs}} t\ \mathrm{SFR}(t)\ dt}{\int_{0}^{t_\mathrm{obs}} \mathrm{SFR}(t)\ dt}.
\end{equation}

\noindent This parameter gives an indication of the epoch at which the stellar masses of galaxies were assembled, which should ultimately agree with the measured redshift evolution of the cosmic SFRD (see Section \ref{sect:madau}).

In Section \ref{subsect:priors_z0} we report the priors imposed on sSFR and mass-weighted formation time by our fiducial models at $z=0$. Then, in Section \ref{subsect:priors_ppds}, we consider the effects of varying the prior probability densities for individual model parameters. Finally, in Section \ref{subsect:priors_higherz}, we explore the effects of moving to higher observed redshifts.

\subsection{Fiducial models at redshift zero}\label{subsect:priors_z0}

We first sample the fiducial prior distributions shown in Table \ref{table:priors} at a fixed redshift of $z = 0$. Fig. \ref{fig:fiducial_priors} shows the priors imposed on sSFR, $t_\mathrm{MW}$ and SFH shape by the four parametric models.

The far-left panels show the priors imposed on sSFR, with the consensus $z=0$ SFMS reported by \cite{Speagle2014} shown as a solid blue line (the shaded region represents a scatter of 0.3 dex). It can be seen that the sSFR priors imposed by all four models are strongly peaked around sSFR $\sim 10^{-10}$ yr$^{-1}$, with tails out to lower sSFRs. A limit of sSFR $\leq 10^{-8}$ yr$^{-1}$ is imposed by the use of a SFR timescale of 100 Myr. 

Given that the SFRs of galaxies are poorly constrained by the observed SED (see Section \ref{sect:intro}), the fact we observe different sSFR priors for different models is one of the main reasons that different locations are observed for the SFMS in different studies (e.g. \citealt{Speagle2014}; \citealt{Pacifici2015}). However, the strongly peaked (and hence informative) nature of these prior distributions also suggests another possibility: that galaxy SFRs measured assuming these parametric SFH models could be driven towards a narrow range of sSFR values by the SFH prior. This could cause a SFMS to be observed even when the data does not have the necessary constraining power to infer reliable SFRs, or artificially tighten an otherwise less-well-defined SFMS.

The centre-left panels show the priors imposed on mass-weighted formation time, with the mass-weighted formation time for the stellar population of the Universe at $z=0$, as calculated from the \cite{Madau2014} SFRD curve, shown by a solid green line. It can be seen that all of the models we consider favour larger $t_\mathrm{MW}$ than \cite{Madau2014}, with the tau and delayed models in particular favouring young galaxy stellar populations. The lognormal and DPL models produce broader priors on $t_\mathrm{MW}$, but still favour later times than the epoch at which it is known that galaxies assembled the majority of their stellar masses. The right panels show the priors imposed on SFH shape by dividing out the dependency on $M_\mathrm{formed}$. The \cite{Madau2014} curve is also shown, clearly demonstrating that these fiducial priors favour later stellar-mass assembly. 

In response to issues of this nature, there has been a  recent rapid increase in the diversity and complexity of parametric SFH models in use in the literature (e.g. \citealt{Ciesla2017}; \citealt{Glazebrook2017}; \citealt{Merlin2018}; \citealt{Carnall2018}; \citealt{Schreiber2018}). This increasing diversity in applied methodologies necessitates a firm basis for comparisons between different methods.

The method presented in this section (sampling SFHs from the prior probability distributions on model parameters, then using these to construct the priors on the parameters of interest) is applicable to any kind of SFH model. We therefore suggest that these tests should always be conducted when a new model is used, in order to understand the priors being imposed, and to facilitate comparisons with other methods.

This method illustrates the power of the fully Bayesian methodology being pioneered in SED fitting analyses. To give an example of how this could be used to refine a SFH model, a more physical prior could be motivated by the ansatz that the shape prior should resemble the overall cosmic SFRD (e.g. \citealt{Gladders2013}). In this case we would wish to find a prior which rises rapidly to a peak around $z=2$ on the right panels of Fig. \ref{fig:fiducial_priors} and then declines more slowly, with $t_\mathrm{MW}$ peaked around the green line on the centre-left panels (see Appendix \ref{lognorm_dpl_ppds}). Another option would be to design a prior centred on the SFH of a galaxy which follows the SFMS throughout its evolution (e.g. \citealt{Ciesla2017}). If no compelling argument can be made for a prior belief that the distribution of galaxy sSFRs at $z=0$ is  strongly peaked at $\sim 10^{-10}$ yr$^{-1}$, then a less informative prior on sSFR would also be desirable.

\begin{figure}
	\includegraphics[width=\columnwidth]{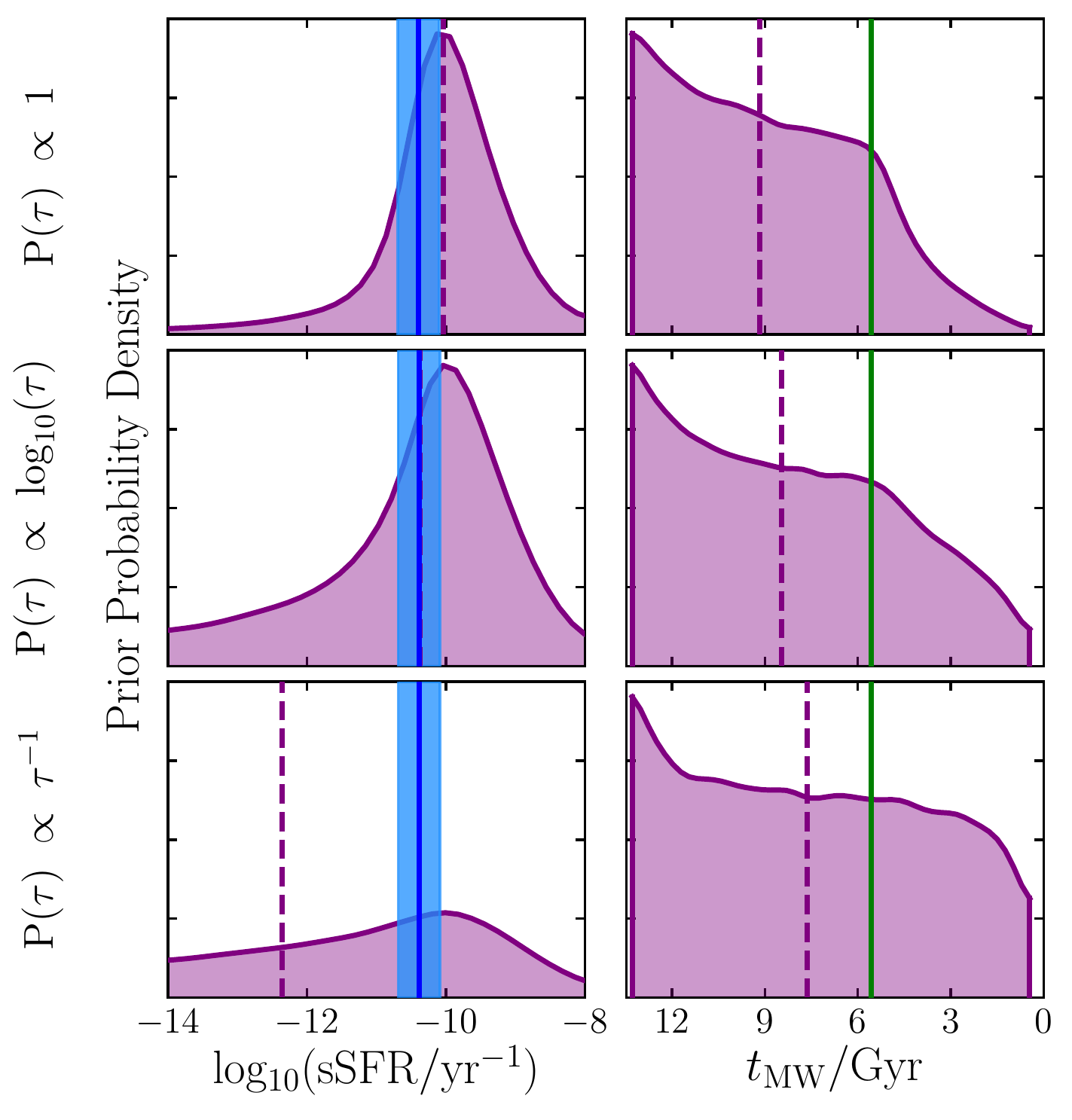}
    \caption{Priors imposed on sSFR and mass-weighted formation time by the exponentially declining SFH parameterisation (Equation \ref{eqn:tau}) under the assumption of different prior probability densities for the $\tau$ parameter. Other priors are the same as listed in Table \ref{table:priors}. Vertical lines are as in Fig. \protect \ref{fig:fiducial_priors}.}\label{fig:alt_tau_priors}
\end{figure}

\begin{figure}
	\includegraphics[width=\columnwidth]{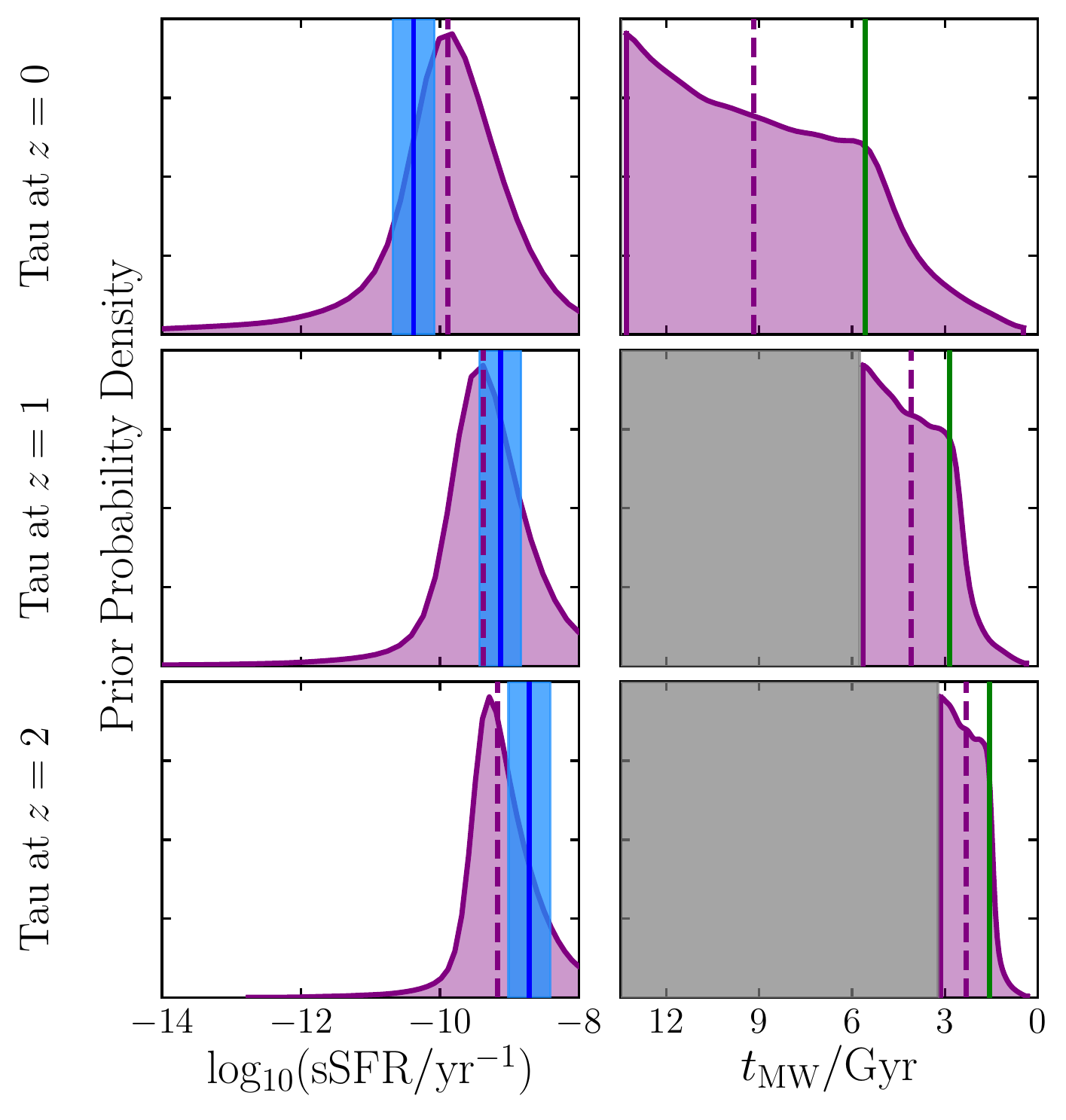}
    \caption{Priors imposed on sSFR and mass-weighted formation time by the exponentially declining SFH parameterisation (Equation \ref{eqn:tau}) with the priors shown in Table \ref{table:priors} at different observed redshifts (different $T_0$ priors). Vertical lines are as described in the Fig. \protect \ref{fig:fiducial_priors} caption at the redshifts shown.}\label{fig:priors_highz}
\end{figure}

\subsection{The effects of changing the prior probability densities assumed for model parameters}\label{subsect:priors_ppds}

The discussion of Section \ref{subsect:priors_z0} refers only to one set of fiducial prior probability densities for the parameters of the four SFH models, chosen to be typical of applications of these models in the literature. Changing any of these will affect the priors shown in Fig. \ref{fig:fiducial_priors}, with a variety of different combinations possible.

Whilst \textsc{Bagpipes} can be used to obtain priors on physical parameters for any of these combinations, we here restrict our discussion to two cases of particular interest involving the prior probability density assumed for the $\tau$ parameter of the exponentially declining SFH model. These provide representative examples from which the magnitude of the effects of such changes can be seen. Another example will be provided in Section \ref{subsect:priors_higherz}, where we will consider changing the observed redshift. This is effectively a specific case of changing the prior on the $T_0$ parameter. An expanded discussion of the prior probability densities for the lognormal and DPL models is also provided in Appendix \ref{lognorm_dpl_ppds}.

An alternative parameterisation of the tau model to that given in Equation \ref{eqn:tau} is $\mathrm{SFR}(t)\propto e^{-\gamma (t-T_0)}$, where $\gamma = \tau^{-1}$ (e.g. \citealt{Pacifici2015, Salim2016}). A uniform prior applied to $\gamma$ corresponds to a prior probability density on $\tau$ of $P(\tau) \propto \tau^{-1}$, as opposed to a uniform prior on $\tau$. The use of a logarithmic prior on $\tau$ is also often discussed. We therefore test two alternative prior probability densities for $\tau$, $P(\tau) \propto \mathrm{log}_{10}(\tau)$ and $P(\tau) \propto \tau^{-1}$ whilst maintaining the ranges and priors for $T_0$ and $M_\mathrm{formed}$ quoted in Table \ref{table:priors}.

Fig. \ref{fig:alt_tau_priors} shows the priors on sSFR and mass-weighted formation time for the two alternative $\tau$ priors, as well as the original uniform prior from Fig. \ref{fig:fiducial_priors} for reference. It can be seen that both of these alternative priors broaden the prior on sSFR towards lower values, with the distribution under the assumption of $P(\tau) \propto \tau^{-1}$ being relatively flat. Additionally, both of these priors extend the distribution of mass-weighted formation times further towards earlier times, again with the assumption of $P(\tau) \propto \tau^{-1}$ producing a relatively flat prior on $t_\mathrm{MW}$. 

We conclude, by a comparison of Figures \ref{fig:fiducial_priors} and \ref{fig:alt_tau_priors}, that changing the prior probability densities assumed for SFH model parameters can produce significant changes in the priors imposed on the parameters of interest, of the same order of magnitude as changing between different SFH parameterisations. We also conclude that when using the exponentially declining SFH model, a prior probability density for the $\tau$ parameter of  of $P(\tau) \propto \tau^{-1}$ produces a less informative prior on parameters of interest than the more common uniform $\tau$ prior, $P(\tau) \propto 1$.

\subsection{The effects of changing the observed redshift}\label{subsect:priors_higherz}

We have so far restricted our discussion to the priors imposed at $z=0$. It is also interesting to consider if and how these priors change as a function of observed redshift. In particular, given our conclusion in Section \ref{subsect:priors_z0} that the priors imposed could bias results from galaxy SED fitting in favour of a tight SFMS within a narrow range in sSFR, it is interesting to consider whether the redshift evolution of the prior on sSFR matches the redshift evolution of the SFMS.

Fig. \ref{fig:priors_highz} shows the priors imposed on sSFR and mass-weighted formation time by the fiducial tau model as described in Table \ref{table:priors} at a range of observed redshifts, chosen to span the cosmic time interval between the epoch of peak star-formation and the present day. Increasing the redshift effectively changes the prior on the $T_0$ parameter by reducing $t_\mathrm{obs}$, the upper limit. This discussion is therefore simply a specific case of changing prior probability densities, as discussed more generally in Section \ref{subsect:priors_ppds}.

It can be seen that the prior on mass-weighted formation time retains the same shape and retains a bias towards later formation than the $t_\mathrm{MW}$ values calculated from the \cite{Madau2014} SFRD curve.

The sSFR priors imposed by this model can be seen to be a strong function of redshift, with the prior evolving towards sharper peaks at higher sSFRs with increasing redshift. This evolution is in the same sense as the SFMS, although the evolution is weaker than for the SFMS of \cite{Speagle2014} at fixed stellar mass.

It is informative to compare the results shown in Fig. \ref{fig:priors_highz} with those of \cite{Ciesla2017}. The left hand panel of their fig. 8 shows that, even at $z=0$, the peak of the sSFR prior for their tau model falls below the SFMS at our fiducial stellar mass of $M_* = 10^{10.5}\mathrm{M_\odot}$. This contrasts with the top two panels of Fig. \ref{fig:priors_highz}, where the prior peaks above the SFMS. This difference is a consequence of different lower limits placed on the $\tau$ parameter: \cite{Ciesla2017} use 1 Myr, as opposed to our limit of 300 Myr.

We conclude that the use of standard parametric SFH models both has the potential to push galaxies towards a narrow range of sSFR values, and that this range of sSFR values evolves with observed redshift in the same sense as the SFMS. Clearly, therefore, an important consideration when designing galaxy SED fitting analyses is to avoid unintentionally biasing results towards recovery of a tight SFMS by the use of a strong prior on sSFR where the observational data being fitted may not be constraining enough to reliably infer SFRs.

\begin{figure}
	\includegraphics[width=\columnwidth]{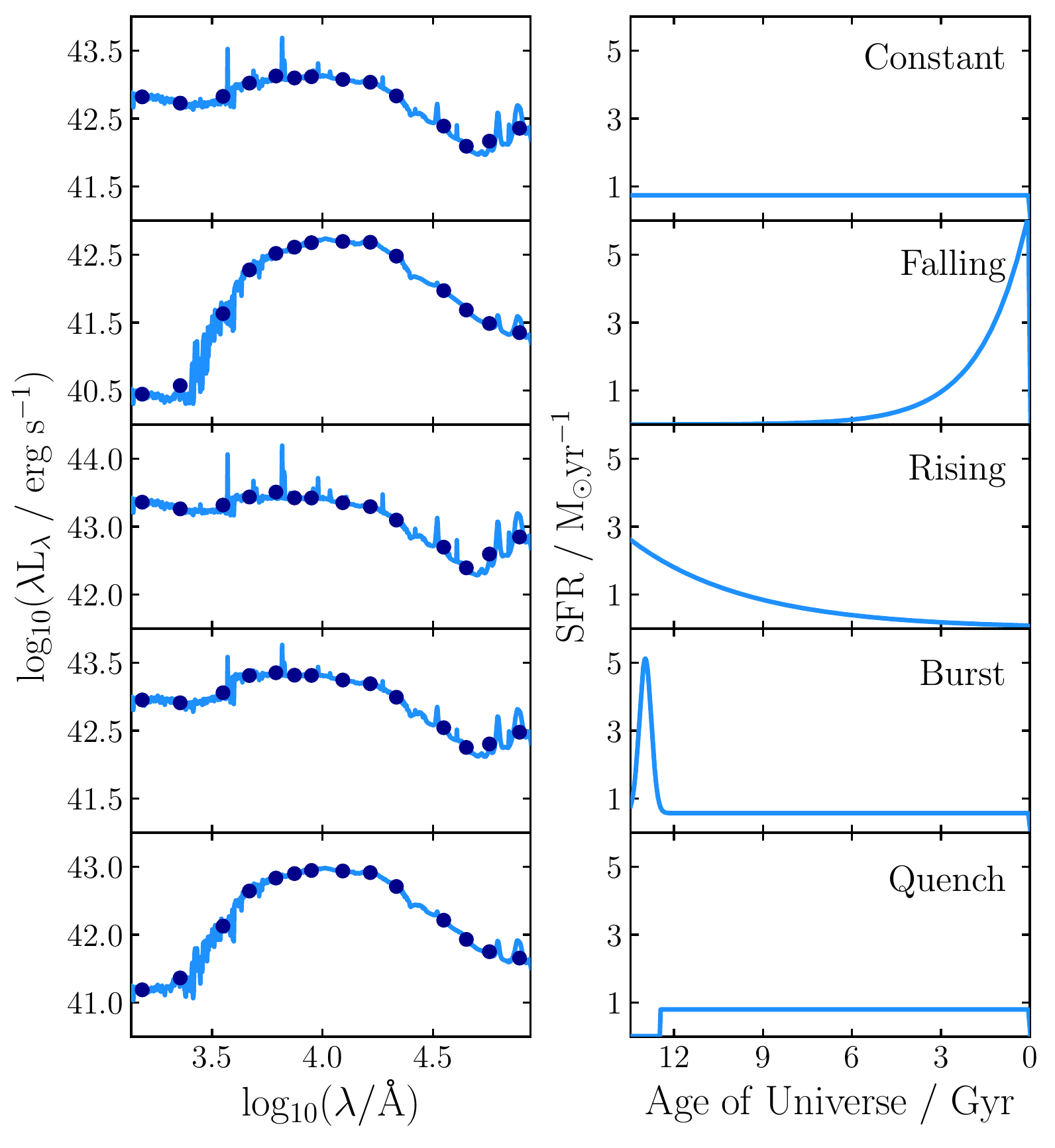}
    \caption{SEDs (left) and SFHs (right) for each of the mock galaxies introduced in Section \protect \ref{subsect:mocks_making}. Photometric fluxes derived from these spectra are shown as circles on the left for each of the following filters: \textit{GALEX FUV/NUV}, SDSS \textit{ugriz}, 2MASS \textit{JHKs} and \textit{Spitzer}/IRAC channels $1-4$.}\label{fig:mocks}
\end{figure}

\section{Testing parametric models with mock observations}\label{sect:mocks}

In Section \ref{sect:priors} we have discussed how, in the absence of strongly constraining data, the priors imposed by the SFH model on physical parameters can affect the results obtained by assigning greater prior weights to SFHs of certain shapes. However, even in the high-SNR regime, the limited range of shapes that a given parametric SFH model is capable of reproducing can introduce biases into physical parameter estimates. To understand these biases it is necessary to examine some test cases involving high-SNR data.

In this section we test the abilities of the parametric SFH models introduced in Section \ref{sect:models} to recover galaxy physical parameters from mock high-SNR broad-band photometric data. In Section \ref{subsect:mocks_making} we describe our mock catalogue, and the generation and fitting of mock photometry with \textsc{Bagpipes}. In Section \ref{subsect:mocks_recovery} we discuss the biases introduced by the use of parametric SFH models to fit these mock data. In Section \ref{subsect:mocks_comparison} we compare the quality of the fits obtained using different parametric SFH models. Finally, in Section \ref{subsect:mocks_information} we consider the implications of our results for the use of broad-band photometric observations as a tool for learning about galaxy SFHs.

\begin{figure*}
	\includegraphics[width=\textwidth]{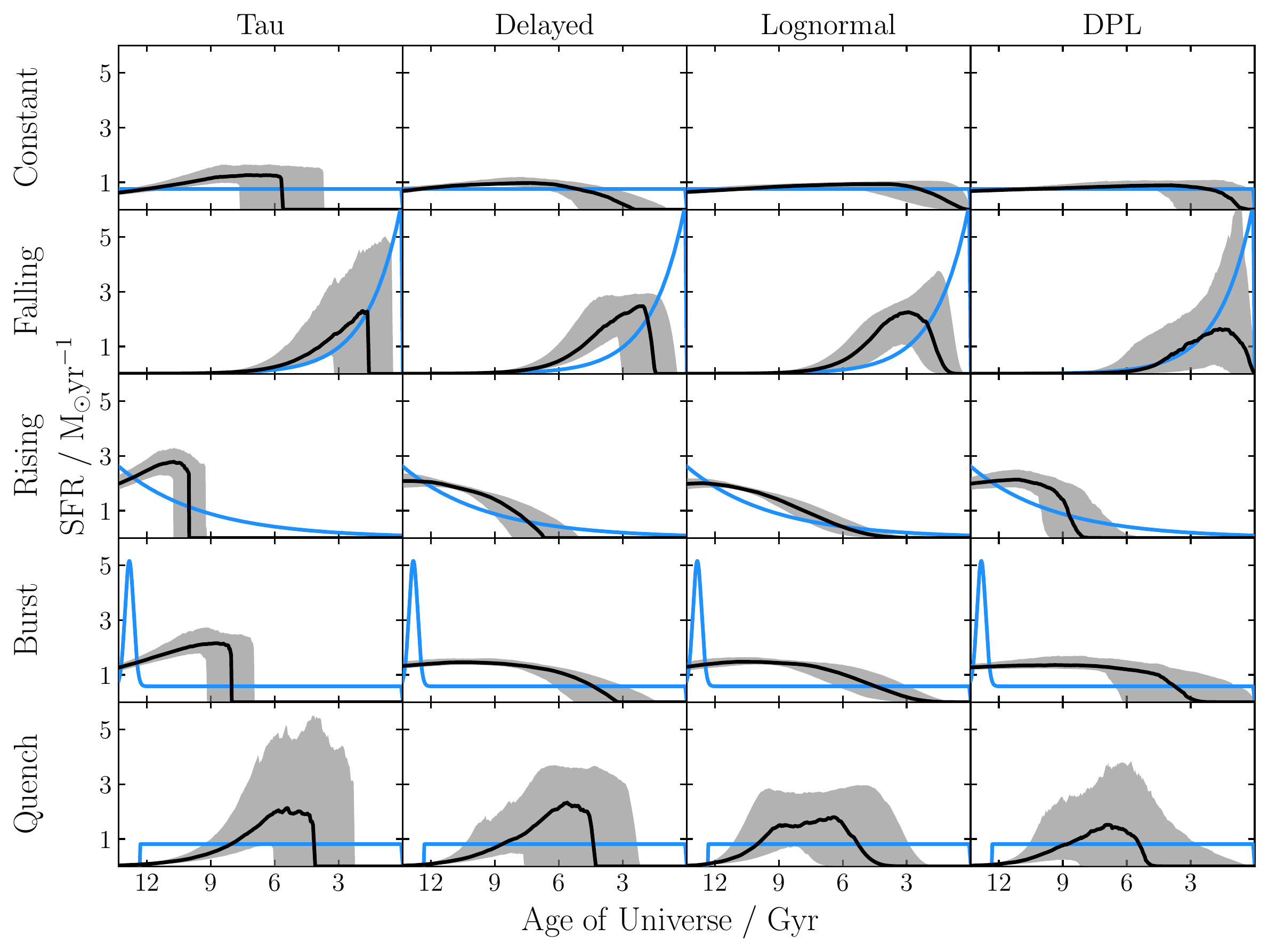}
    \caption{Recovery of SFHs from the mock data described in Section \ref{subsect:mocks_making} using the four parametric SFH models described in Section \ref{sect:models} and Table \ref{table:priors}. Different rows show different mock galaxies, different columns show different fitted SFH models. Blue lines are the input mock SFHs, black lines show the posterior medians and the shaded regions show the 16\textsuperscript{th}--84\textsuperscript{th} percentiles.} \label{fig:fitted_sfh}
\end{figure*}

\begin{figure*}
	\includegraphics[width=\textwidth]{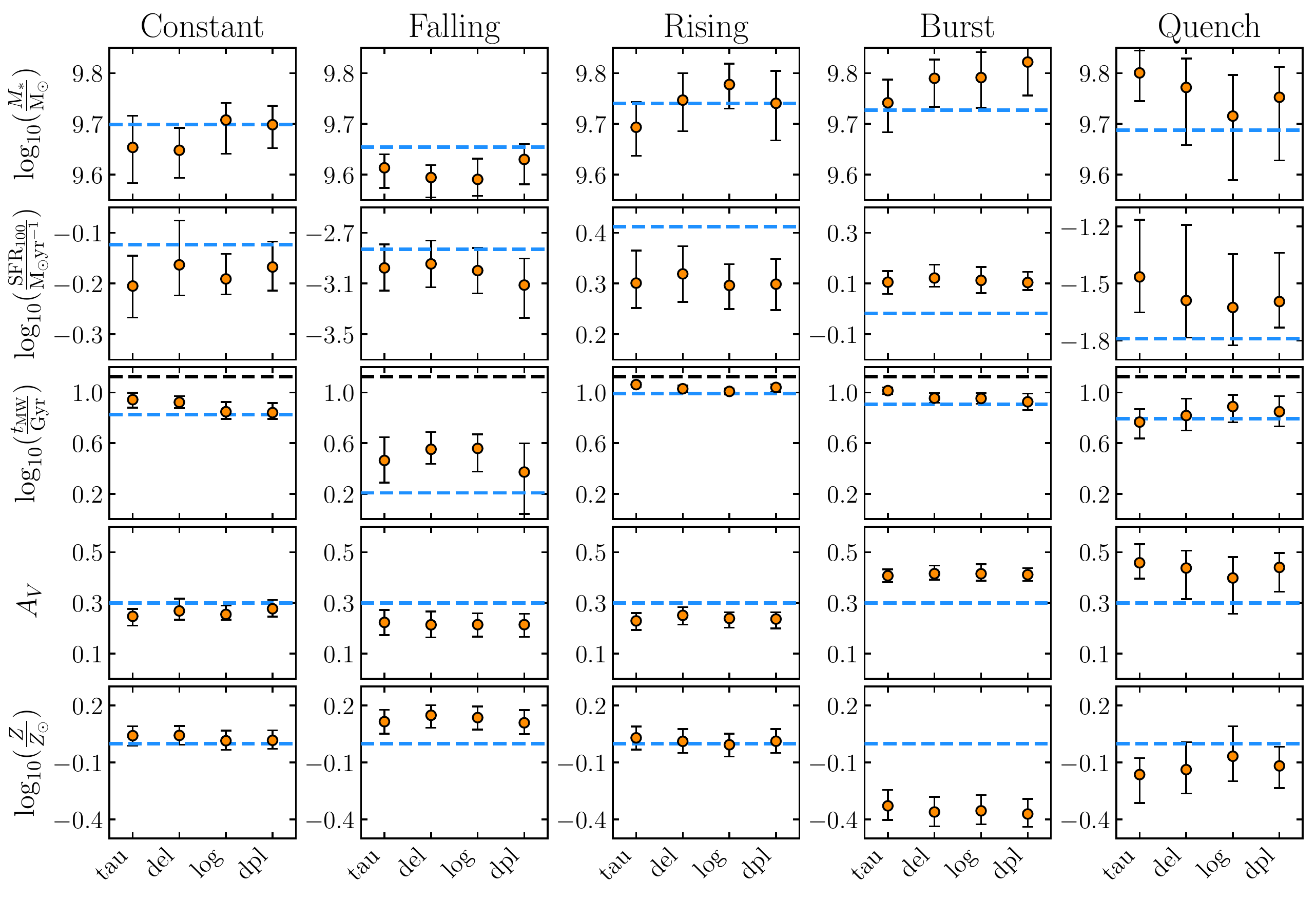}
    \caption{Physical parameter recovery from mock data (see Section \ref{subsect:mocks_making}) using our SFH models (see Section \ref{sect:models} and Table \ref{table:priors}). Posterior median values are shown as orange circles, while the errorbars show the 16\textsuperscript{th}--84\textsuperscript{th} percentiles of the posteriors. Blue dashed lines show the true values, black dashed lines show $t_\mathrm{obs}$. The observed biases result from the parametric SFH models being unable to reproduce the correct SFH shapes (see Fig. \protect \ref{fig:fitted_sfh}) and assigning different prior weights to different SFH shapes.}\label{fig:mocks_1d_post}
\end{figure*}

\begin{figure*}
	\includegraphics[width=\textwidth]{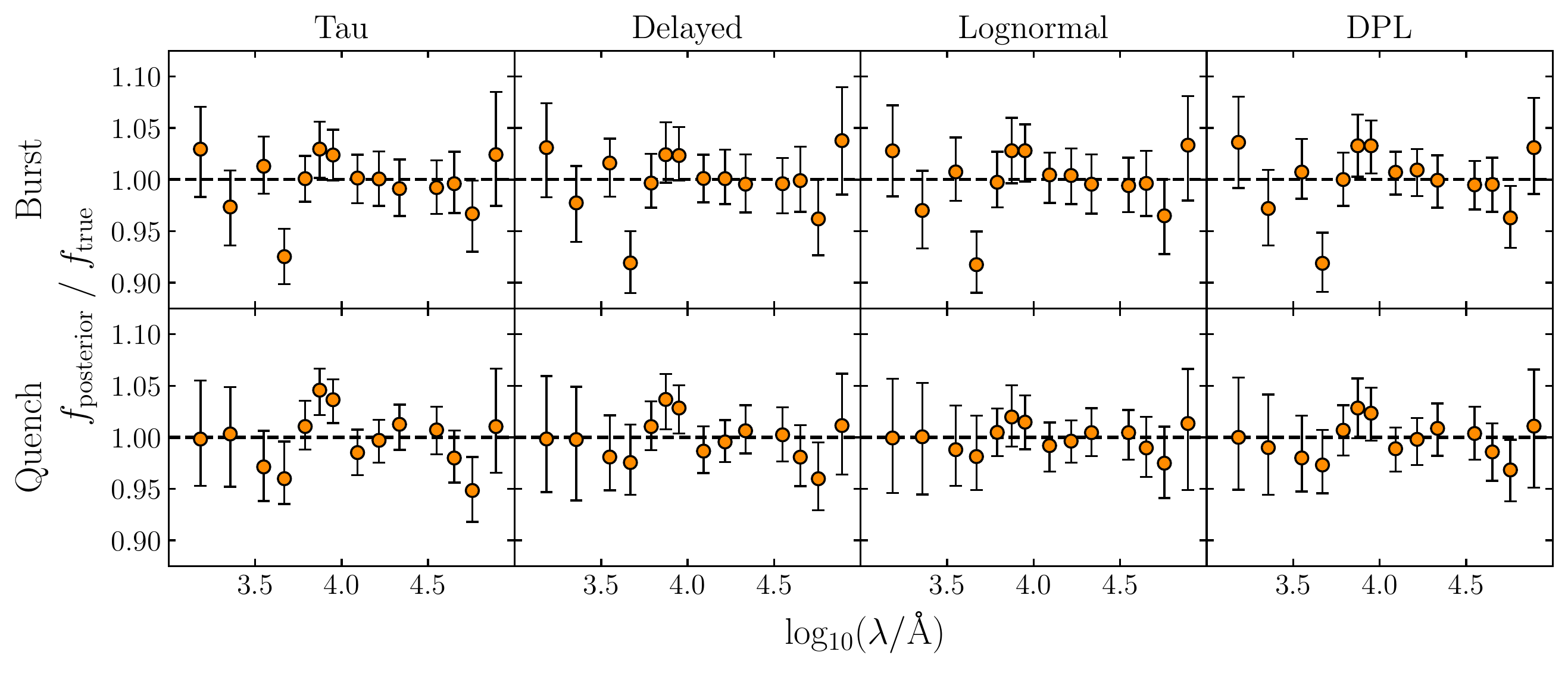}
    \caption{Posterior predictions for photometric observations compared to the input values. The two rows correspond to the recent burst and rapid quench mocks (results for the other three mocks are very similar). The four columns correspond to our parametric SFH models. Posterior medians are shown as orange circles, error bars show the 16\textsuperscript{th}--84\textsuperscript{th} posterior percentiles.}\label{fig:fitted_spec}
\end{figure*}

\subsection{Generating and fitting mock data}\label{subsect:mocks_making}

To facilitate our tests, a set of five mock galaxies at $z=0$ with different SFHs was generated. The SFHs are:

\begin{itemize}

\item \textbf{Constant}: equal SFR from $t=0$ to $t_\mathrm{obs}$.

\item \textbf{Falling}: exponential decline (see Equation \ref{eqn:tau}) with $T_0=0$ and $\tau =  t_\mathrm{obs}/10 = 1.4$ Gyr.

\item \textbf{Rising}: SFR$(t) \propto e^\frac{t}{\tau}$ with $\tau =  t_\mathrm{obs}/4 = 3.4$ Gyr.

\item \textbf{Recent burst}: constant SFR from $t~=~0$ to $t_\mathrm{obs}$ making up 80\% of $M_\mathrm{formed}$ and a Gaussian burst centred 500 Myr before $t_\mathrm{obs}$ with width $\sigma~=~200$ Myr making up the other 20\% of $M_\mathrm{formed}$.

\item \textbf{Sudden quench}: constant star formation from $t=0$ to $t_\mathrm{obs} - 1$ Gyr, then constant star formation at 2\% of the original level from that point to $t_\mathrm{obs}$.
\end{itemize}

Mock photometry was generated with \textsc{Bagpipes} using the methods described in \cite{Carnall2018} for each of the following photometric filters: \textit{GALEX FUV/NUV}, SDSS \textit{ugriz}, 2MASS \textit{JHKs} and \textit{Spitzer}/IRAC channels $1-4$. We assign each photometric flux an uncertainty corresponding to a high SNR of 25. However, we do not perturb the model fluxes by these uncertainties, so as to isolate the effects of the SFH parameterisation. The mock SEDs and SFHs are shown in Fig. \ref{fig:mocks}.

Each model was assigned a total stellar mass formed, $M_\mathrm{formed}=10^{10}$ $\mathrm{M_\odot}$ and a metallicity of $Z = 0.02$ (which we take to be Solar metallicity, $Z_\odot$). An ionization parameter of log$_{10}(U)~=~-3$ was assumed for nebular emission. Dust attenuation with $A_V$ = 0.3 mag was applied using the \cite{Calzetti2000} attenuation curve. Attenuation was assumed to be doubled for stars formed in the last 10 Myr and for nebular emission. All attenuated light was assumed to be re-radiated in the infrared using the dust emission models of \cite{Draine2007}, (recently added to \textsc{Bagpipes}). We assume values of 2 for $Q_\mathrm{pah}$, the percentage of dust mass in polycyclic aromatic hydrocarbons, 1 for $U_\mathrm{min}$, the minimum starlight intensity to which the dust is exposed, and 0.01 for $\gamma_\mathrm{e}$, the fraction of the incident starlight at $U_\mathrm{min}$.

Our mock catalogue was chosen to span a wide range of scenarios for the formation of galaxies. However it should be noted that, by constructing this catalogue, we are expressing prior beliefs about the SFHs of real galaxies. The results of Sections \ref{subsect:mocks_recovery} and \ref{subsect:mocks_comparison} are necessarily dependent on these prior beliefs (see Section \ref{subsect:mocks_information}). The same mock catalogue is used in our companion paper \citep{Leja2018a}, however it should be noted that a number of different assumptions are made by \textsc{Bagpipes} and \textsc{Prospector}, in particular for the stellar/nebular emission and dust attenuation models.

Each of the four parametric models introduced in Section \ref{sect:models} was fitted to each set of mock photometry assuming the priors given in Table \ref{table:priors} at a fixed redshift of $z=0$. The other free parameters are $A_V$ for the \cite{Calzetti2000} dust attenuation law, to which we assign a uniform prior between 0 and 4 mag, metallicity, $Z$, to which we assign a logarithmic prior over the range $0.2 < Z/Z_\odot < 2$, and the dust emission parameters, to which we assign the same priors as \cite{Leja2017}, except that we do not extrapolate $Q_\mathrm{pah}$ beyond a maximum value of 4.58.

\subsection{Recovery of SFHs and physical parameters}\label{subsect:mocks_recovery}

Fig. \ref{fig:fitted_sfh} shows the posterior SFHs obtained by the process described in Section \ref{subsect:mocks_making} compared to the input mock SFHs. Fig. \ref{fig:mocks_1d_post} shows the posterior constraints obtained for a number of physical parameters compared to their true values. It can be seen that the quality of the recovered posterior SFHs and physical parameters is a strong function of the shape of the input mock SFH and a weaker, but still significant, function of the parametric SFH model which is fitted. This finding is supported by Fig. \ref{fig:mocks_1d_post}, where it can be seen that often all four models return consistent values for physical parameters which are in strong tension with the true value.

The SFHs of the first three ``simple" mocks (constant, falling, rising) are relatively well recovered by all of the SFH models, with the exception of the tau model, which predictably struggles to recover the constant and rising SFHs. It is also interesting to highlight the case of the tau model fitting the falling mock, as in this case the true SFH exists within the prior. It can be seen from Fig. \ref{fig:fitted_sfh} that, whilst the tau model recovers the shape of the input SFH well, the recovery is not perfect as might be expected. This is due to the non-Gaussian shape of the posterior distribution, in this case for the $T_0$ and $\tau$ parameters.

This is a common consequence of poorly constraining data, and means that the most probable (maximum a posteriori) and posterior median parameter estimates may be offset. In this case, because the model fluxes have not been perturbed, the most probable model is the input model. However, when dealing with real observational uncertainties and non-Gaussian posteriors, the maximum a posteriori parameter estimates are often poorly representative of the posterior.

For the ``simple'' mocks, the stellar mass, SFR, dust attenuation and metallicity posteriors can be seen from Fig. \ref{fig:mocks_1d_post} to typically fall within $1-2\sigma$ of the true values, with biases of the order of $\sim0.1$ dex for stellar mass and $\sim0.2$ dex for SFR. The posterior mass-weighted formation times are more strongly in tension with the input values (up to $\sim4\sigma$), and are typically overestimates of the true values. These biases are due to all four models, to some extent, failing to reproduce the early-time evolution of these mock SFHs.

By contrast, all four SFH models struggle to reproduce the more complex shapes of the recent burst and sudden quench models. More significant biases in the recovery of physical parameters can be seen in these cases, up to $0.3$ dex in SFR, 0.2 mag in dust attenuation, and 0.3 dex in metallicity. For the burst mock in particular, the posterior probability distributions strongly exclude the true values for all parameters except $M_*$. The posterior SFHs shown in Fig. \ref{fig:fitted_sfh} appear unable to simultaneously reproduce both the overall shapes of these mock SFHs and their rapid variability at late times, with the shapes fitted representing a compromise between the two. The strong biases in the recovered dust and metallicity values are a consequence of the age-metallicity-dust degeneracy, which allows the mock photometry to be well matched even whilst fitting radically different SFHs.

Based upon these results, we conclude that the choice of parametric SFH model has the potential to significantly affect the physical parameter estimates obtained when fitting high-SNR broad-band photometric observations, in particular by at least 0.1, 0.3 and 0.2 dex for stellar mass, SFR, mass-weighted formation time respectively. Our finding of $\sim0.1$ dex variations in stellar-mass measurements is in good agreement with similar analyses in the literature (e.g. \citealt{Pacifici2015, Mobasher2015, Iyer2017}). For SFRs however, both \cite{Pacifici2015} and \cite{Iyer2017} find a wider range of variations, suggesting our limited library of mocks does not encompass the most pathological cases.

However, often all four of our models return consistent posterior estimates for physical parameters which are strongly biased from the input values. This suggests that the true shape of the SFH is the most important factor in whether parametric models can return unbiased physical parameter values, with even a simple mock catalogue containing SFHs which are not well described by any of our parametric models. The parametric models are least reliable for galaxies which experience recent, rapid changes in their SFRs. This limitation can be improved upon by the use of non-parametric models \citep{Leja2018a}.

\subsection{Comparisons between SFH models}\label{subsect:mocks_comparison}

Given the biases we have discussed in Section \ref{subsect:mocks_recovery}, it is interesting to consider whether there is a case, based on our results, for using one of these parametric models in preference to others. In the ideal case, we would be able to identify one parametric model which reliably returns smaller biases in physical parameter estimates than the others and recommend this for general use. Failing this, we would hope to be able to identify the model which returns the least biased physical parameter estimates for a specific object (without knowledge of the true values) by assessing the relative quality of the fits. It should again be stressed, as in Section \ref{subsect:mocks_making}, that any conclusions as to which parametric SFH models are more preferred depend on the prior beliefs about real galaxy SFHs we expressed when building our mock galaxy catalogue.

Considering the relative biases returned by the use of different models, as discussed in Section \ref{subsect:mocks_recovery}, often our different SFH models return consistent parameter estimates, all of which are biased with respect to the input values. In cases where the relative biases differ, the tau model generally returns more highly biased physical parameter estimates, however there is little distinction between the biases obtained when using the other three parametric models. We therefore conclude that none of the delayed, lognormal or DPL models can be said to reliably produce less-biased physical parameter estimates than any other when fitting high-SNR photometric data.

\begin{figure*}
	\includegraphics[width=\textwidth]{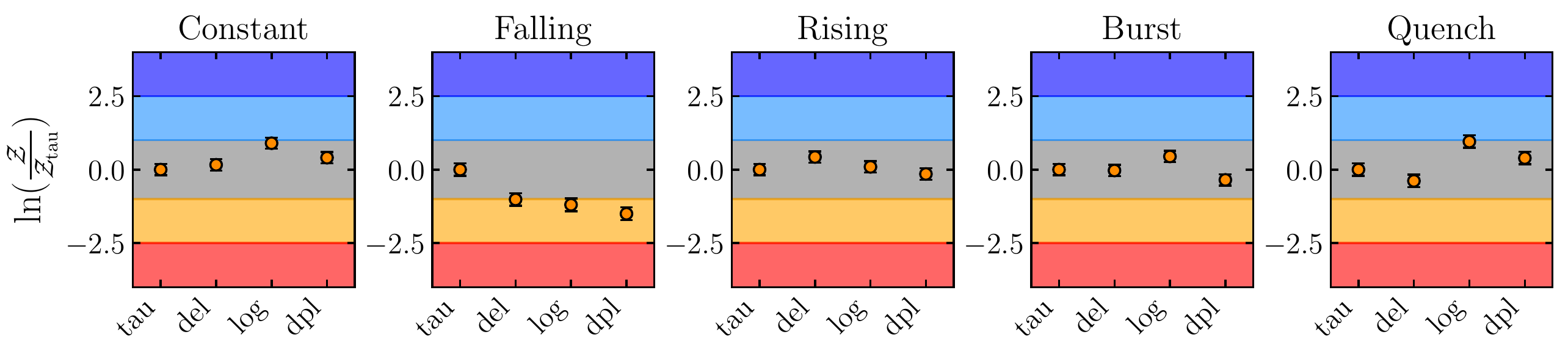}
    \caption{Bayesian evidence plots (e.g. \citealt{Trotta2008}) for each SFH model fitted to each mock, relative to the tau model. The coloured stripes denote (from top to bottom): a strong preference for (dark blue), a weak preference for (light blue), no preference for (grey), a weak preference against (orange) and a strong preference against (red) the model compared to the tau model. It can be seen that for none of these mocks is there a strong preference for any one of the SFH models above the others.}\label{fig:evidence}
\end{figure*}

We therefore move on to consider which models produce the best fits to our mock data. Fig. \ref{fig:fitted_spec} shows the residuals between our posterior predictions for photometry and our input mock data. Results are shown for each of our four SFH models fitted to the recent burst and sudden quench mocks (results for the other three mocks are very similar). It can be seen that all of the models produce posterior distributions which appear to be acceptable fits to the mock photometry. Furthermore, it is impossible to distinguish by eye between the posteriors obtained by fitting each of the different models.

In order to quantify this we consider the Bayesian evidence for each model fitted to each set of mock photometry. Evidence is analogous in a Bayesian framework to the more commonly used minimum reduced chi-squared value, and is often used to discriminate between models (e.g. \citealt{Trotta2008, Salmon2016}). Fig. \ref{fig:evidence} shows the evidence for each model fitted to each set of mock photometry relative to the tau model. The coloured stripes represent (from top to bottom) strong evidence for, weak evidence for, no evidence for, weak evidence against and strong evidence against one model compared to another.

It can be seen from Fig. \ref{fig:evidence} that none of our parametric models is strongly favoured or disfavoured compared to the tau model for any of the mocks. Even for the falling mock, where the input model is within the tau-model prior, only a very weak preference is visible. This means that it is not possible, in general, to identify which SFH model produces the least biased physical parameter estimates for individual objects by assessing the ``goodness of fit'' to broad-band photometry. A specific case was recently demonstrated by \cite{Belli2018}, who show that it is not possible to distinguish between parametric SFH models for a sample of quiescent galaxies.

\subsection{Broad-band photometry as a tool for understanding galaxy SFHs}\label{subsect:mocks_information}

In Section \ref{subsect:mocks_comparison} we have demonstrated that, even though we observe from Fig. \ref{fig:mocks_1d_post} that the tau model produces more highly biased physical parameter estimates for our mock catalogue, this is not associated with a worse quality of fit. This means that, for the case in which we have no knowledge of the true physical parameter values, we have no basis for deciding which parametric SFH model produces less-biased physical parameter estimates.

Our conclusion that the tau model produces more highly biased physical parameter estimates in Section \ref{subsect:mocks_comparison} is a consequence of the prior beliefs about real galaxy SFHs we expressed when constructing our mock catalogue in Section \ref{subsect:mocks_making}. If all of the mocks in our catalogue were similar to the falling mock, we would conclude that the tau model describes galaxy SFHs equally as well or better than the other three models. 

This leads us to an important conclusion: high-SNR broad-band photometry cannot, in general, constrain prior beliefs about which parametric SFH models are most appropriate for describing the SFHs of real galaxies. Hypotheses such as galaxy SFHs being well described by the lognormal function \citep{Gladders2013} cannot be proven or disproven using broad-band photometric observations. The physical parameter inferences made are therefore necessarily dependent on prior beliefs, to at least the levels reported in Section \ref{subsect:mocks_recovery}. 

There are three possible responses to this conclusion. Firstly, one can perform a more sophisticated version of the analysis presented in Sections \ref{subsect:mocks_making} and \ref{subsect:mocks_recovery}, by generating a catalogue of mock data which represents a set of prior beliefs about galaxy formation (e.g. \citealt{Buat2014}; \citealt{Ciesla2015}), then fitting those mocks with different parametric SFH models to decide which is the most appropriate for representing those prior beliefs.

A popular choice is to use catalogues of mock observations drawn from simulations of galaxy formation (e.g. \citealt{Pacifici2015, Diemer2017, Carnall2018}). However, the lack of flexibility in parametric SFH models means that the process of selecting an appropriate model usually involves a significant amount of trial and error, with no clear physical link between the chosen model and the prior beliefs expressed. Additionally, the use of prior beliefs drawn from simulations renders comparisons between observational results and simulation outputs of questionable value.

This brings us to the second option: the use of non-parametric SFH models. Because of the increased flexibility of these models, it is far easier to encode more specific prior beliefs about the shapes of galaxy SFHs into the model, with or without the use of mock observations drawn from simulations. This option is the subject of a companion paper \citep{Leja2018a}.

The final option is the analysis of higher quality observational data, such as high-SNR continuum spectroscopy, which has been shown to be more strongly constraining on galaxy SFHs (e.g. \citealt{Gallazzi2005, Gallazzi2008, Ocvirk2006, Pacifici2012, Thomas2017}; Carnall et al. in preparation). These kinds of analyses are extremely promising, as they have the potential to demonstrate a clear preference in favour of one model for the SFHs of galaxies over others, providing deeper insights into the physics driving the assembly of stellar mass in galaxies.

\begin{table}
  \caption{A comparison of $z\sim0.05$ SFRD and SMD estimates. The first four rows show our estimates using different SFH models to fit the GAMA sample (see Section \ref{subsect:madau_data}). We also show several results from the literature (converted to our IMF where necessary). M14 is \cite{Madau2014}, W17 is \cite{Wright2017}, and D18 is \cite{Driver2018}. The uncertainties quoted do not include systematic effects.}
  \begin{center}
\begingroup
\setlength{\tabcolsep}{5pt}
\renewcommand{\arraystretch}{1.4}
\begin{tabular}{ccc}
\hline
Model & log$_{10}\bigg(\dfrac{\mathrm{SMD}}{\mathrm{M_\odot\ Mpc^{-3}}}\bigg)$ & log$_{10}\bigg(\dfrac{\mathrm{SFRD}}{\mathrm{M_\odot\ yr^{-1}\ Mpc^{-3}}}\bigg)$\\
\hline
Tau & $8.40^{+0.02}_{-0.03}$ & $-2.041^{+0.002}_{-0.001}$ \\
Delayed & $8.39^{+0.02}_{-0.02}$ & $-2.033^{+0.001}_{-0.001}$ \\
Lognormal & $8.37^{+0.02}_{-0.02}$ & $-2.082^{+0.002}_{-0.001}$ \\
DPL & $8.38^{+0.03}_{-0.03}$ & $-2.088^{+0.001}_{-0.002}$ \\
MD14 & 8.55 & $-1.97$ \\
W17 & $8.35^{+0.06}_{-0.07}$ & -- \\
D18 & $8.30^{+0.01}_{-0.01}$ & $-1.95^{+0.00}_{-0.00}$ \\
 \hline
\end{tabular}
\endgroup
\end{center}
\label{table:smd}
\end{table}

\section{Testing parametric models with observational data}\label{sect:madau}

In Sections \ref{sect:priors} and \ref{sect:mocks} we have demonstrated the effects of different parametric SFH models on results obtained through SED fitting for galaxy stellar masses, SFRs and mass-weighted formation times. As noted in Section \ref{sect:intro}, all measurements of these quantities are dependent, to some extent, on the SFH model. However, it is possible to minimise the impact of the SFH by looking at SFR indicators which are sensitive to star-formation on very short timescales, over which it is safe to assume that star-formation is constant (e.g. \citealt{Kennicutt2012}). By making these observations for representative samples of galaxies across cosmic time, it is possible to infer the redshift evolution of the cosmic SFRD and hence stellar-mass density (SMD) independently of individual galaxy SFHs (e.g. \citealt{Madau2014}).

In this section, we fit a low-redshift, volume-complete sample of galaxies using our parametric SFH models. We then consider whether the results obtained through SED fitting for the redshift evolution of the cosmic SFRD and SMD are consistent with these ``SFH-free" estimates. This provides a valuable additional perspective on the physical motivation for the priors we are imposing by the use of parametric SFH models on stellar masses, SFRs and mass-weighted formation times (e.g. \citealt{Heavens2004, Ocvirk2006, Gallazzi2008, Wuyts2011}). We will also compare our results to similar SED fitting analyses in the literature as a check on our \textsc{Bagpipes} fitting methodology.

In Section \ref{subsect:madau_data} we discuss the fitting of our observational sample. In Section \ref{subsect:madau_smd} we consider our stellar-mass measurements by calculating the cosmic SMD. In Section \ref{subsect:madau_sfrs} we consider our SFR measurements by calculating the cosmic SFRD and comparing to independent, H$\alpha$-derived SFRs. Finally, in Section \ref{subsect:madau_shape} we consider our mass-weighted formation time measurements by considering the shape of our inferred SFRD evolution.

\subsection{GAMA data and fitting methodology}\label{subsect:madau_data}

In this section, we use data from the Galaxy and Mass Assembly Survey (GAMA; \citealt{Driver2009, Driver2016, Baldry2018}) DR3 release. In particular, the 21-band aperture-matched catalogue for the equatorial G09, G12 and G15 fields generated using the \textsc{Lambdar} code by \cite{Wright2016}. The catalogue includes far-UV to far-IR photometry from \textit{GALEX}, SDSS, VISTA, \textit{WISE} and \textit{Herschel} over a 180 deg$^{2}$ area.

We define our sample by using the stellar mass estimates of \cite{Taylor2011} to select objects with $M_* > 10^9~\mathrm{M_\odot}$. GAMA is mass-complete down to $10^9~\mathrm{M_\odot}$ at $z \lesssim 0.08$ \citep{Lange2015, Driver2016}, and we therefore select objects with GAMA spectroscopic redshifts in the range $0.05 < z < 0.08$. This results in a sample of 6134 objects. Whilst this is, in reality, a mass-complete rather than volume-complete sample, $\sim95$\% of the stellar mass in the Universe at $z=0$ is in galaxies above this mass limit \citep{Tomczak2014}. We therefore approximate our sample as volume-complete.

Each object is fitted with \textsc{Bagpipes} using each of the four parametric SFH models described in Section \ref{sect:models}, under the same assumptions as described in Section \ref{subsect:mocks_making}. To calculate the inferred cosmic SFRD evolution, for each SFH model we extract 100 posterior draws for the SFH of each object. These are then summed across objects and divided by the comoving volume from which our sample is drawn. We use these curves to calculate the inferred SMD and SFRD at $z\sim0.05$ for each SFH model. These values are reported in Table \ref{table:smd}.

\begin{figure}
	\includegraphics[width=\columnwidth]{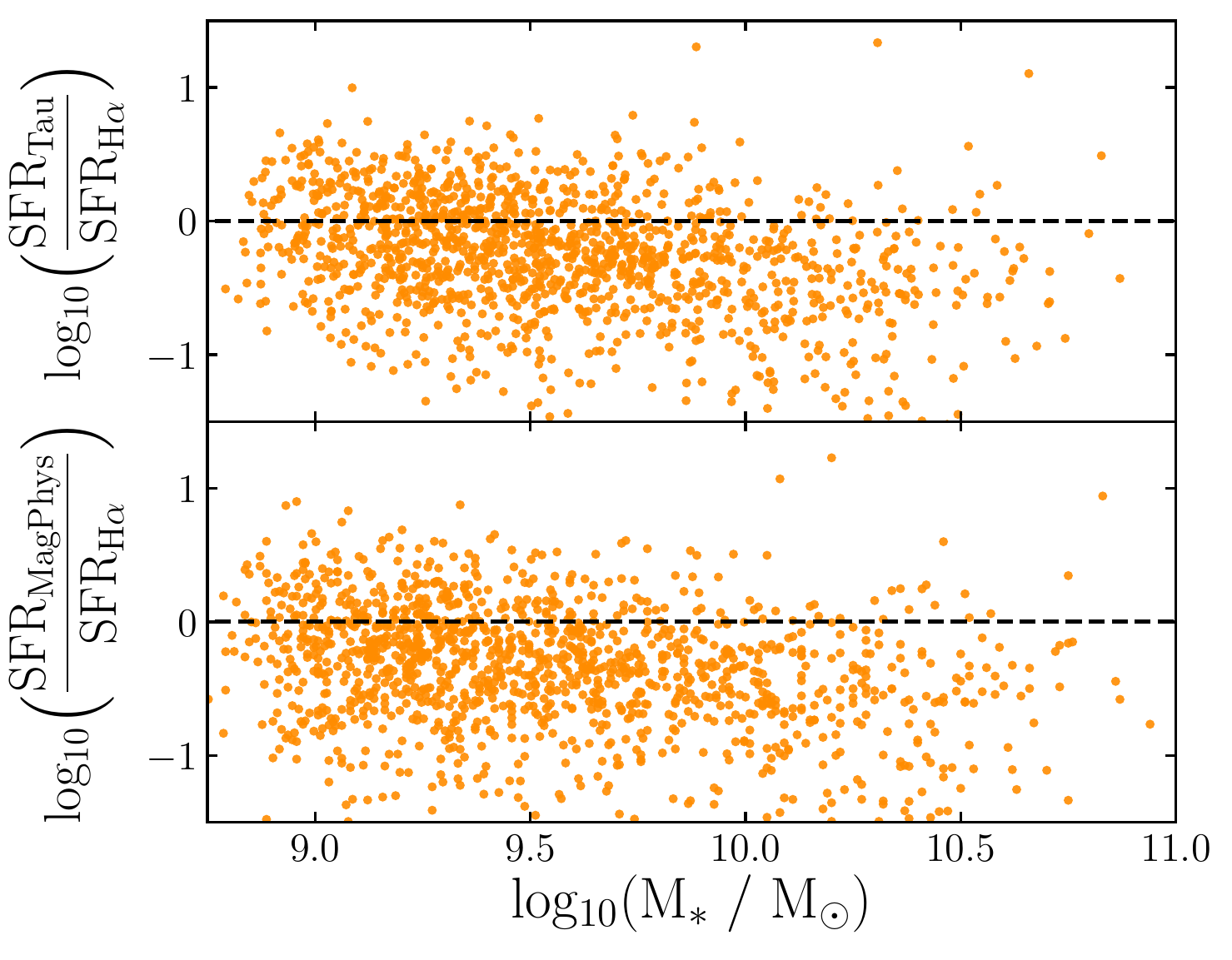}
    \caption{Comparisons between SFRs derived from SED fitting and from dust-corrected H$\alpha$ fluxes using the \cite{Kennicutt2012} relation. The top panel shows SFRs obtained using our tau SFH model; our other models produce similar results. The bottom panel shows SFR estimates from \textsc{MagPhys}, released as part of GAMA DR3.}\label{fig:sfrs}
\end{figure}

\subsection{Inferred stellar masses}\label{subsect:madau_smd}

We first consider the stellar masses we infer for our sample. We evaluate our stellar masses by comparing the cosmic SMD results for each model, shown in Table \ref{table:smd}, to results from the literature. Our results can be seen to be consistent with those of \cite{Wright2017}, derived from the same catalogue using the \textsc{MagPhys} code \citep{daCunha2008}. A similar analysis by \cite{Driver2018} yields a slightly lower value, however all of these results are consistent to well within the $\sim0.3$ dex systematic uncertainties normally assumed for stellar mass measurements (e.g. \citealt{Mobasher2015}). We thus conclude that our stellar mass estimates are consistent with similar analyses in the literature.

A ``SFH-free" estimate of the cosmic SMD can also be obtained by integrating the \cite{Madau2014} SFRD curve, then multiplying by $(1 - R)$, where $R = 0.27$ is the mass-return fraction. A $\sim0.2$ dex offset has been widely observed between this estimate and those from SED fitting analyses (e.g. \citealt{Leja2015}). Our results also reflect this tension, and although this offset is within the systematic uncertainties, the question remains as to which systematic effect is responsible. Our results in Section \ref{subsect:mocks_recovery} suggest that the use of parametric SFH models can lead to systematic offsets in stellar-mass measurements at levels of $\sim0.1$ dex, with the main driver of these offsets being the shape of the true SFH. It is possible therefore that the majority of this offset is due to the biasing effects of parametric SFH models. However this would only be true in the scenario that all galaxy SFHs have true shapes which cause parametric models to underestimate their stellar masses (e.g. similar to our falling mock).

\subsection{Inferred star-formation rates}\label{subsect:madau_sfrs}

We now consider the star-formation rates we infer for our sample. As can be seen from Table \ref{table:smd}, our SFRD results at $z\sim0.05$ fall $\sim0.1$ dex lower than both the SED fitting analysis of \cite{Driver2018} and the SFRD curve of \cite{Madau2014}. However the offset is, again, well within the systematic uncertainties of $\sim0.5$ dex normally assumed for SFR measurements (e.g. \citealt{Pacifici2015}). It is interesting to note that, whereas our SMD measurements reported in Table \ref{table:smd} are consistent, our SFRD measurements are strongly inconsistent with each other. This suggests that SFR measurements are more strongly biased by SFH priors than stellar mass measurements.

We can also assess the quality of the SFRs we derive on an individual basis by comparing our SFRs to those inferred from H$\alpha$ fluxes. GAMA DR3 includes calibrated measurements of H$\alpha$ and H$\beta$ fluxes from GAMA and Sloan Digital Sky Survey (SDSS) spectra. We begin by selecting the 1373 objects of the 6134 in our sample which have SNR $> 5$ in both H$\alpha$ and H$\beta$. We then correct the H$\alpha$ fluxes for dust attenuation using the measured Balmer decrements, following the process outlined in section 3 of \cite{Dominguez2013}. We finally convert the dust-corrected H$\alpha$ fluxes to SFRs using the calibration of \cite{Kennicutt2012}.

Fig. \ref{fig:sfrs} shows our tau-model SFRs compared to those derived from H$\alpha$ as a function of our inferred stellar masses. SED-derived SFRs, calculated using \textsc{MagPhys}, were also released as part of GAMA DR3. These are also shown compared to H$\alpha$ on Fig. \ref{fig:sfrs}. It can be seen that both sets of results agree well with H$\alpha$ at lower masses, however SFRs at progressively higher masses are increasingly under-predicted with respect to H$\alpha$.

It should be noted that the scatter observed in Fig. \ref{fig:sfrs} is partially due to variations in galaxy SFRs on the very short timescales to which H$\alpha$ is sensitive. Our parametric models are not capable of reproducing variations on such short timescales, however several approaches have been demonstrated which allow parametric models to reproduce this behaviour, such as the addition of bursts of star-formation (e.g. \citealt{daCunha2008}), and resampling the average SFR over the last 10 Myr from a separate distribution (e.g. \citealt{Pacifici2016}).

The change we observe in the mean SFR offset with stellar mass is consistent with our result from Section \ref{subsect:mocks_recovery}, that biases in SFRs inferred using parametric SFH models are a strong function of the true SFH shape. Lower-mass galaxies are known to form their stars later in cosmic history, and are likely to have SFHs consistent with our constant and/or rising mocks, for which we recover either unbiased or slightly underestimated SFRs.

Conversely, the significant underestimation of SFRs with respect to H$\alpha$ at higher masses in Fig. \ref{fig:sfrs} is not consistent with the SFR offsets observed for any of our mocks in Figure \ref{fig:mocks_1d_post}. As in Section \ref{subsect:mocks_recovery}, this again suggests that our mock catalogue does not encompass the most pathological cases. However it should also be noted that the dustier nature of higher mass galaxies (e.g. \citealt{Garn2010}) makes H$\alpha$ a less-reliable SFR indicator.

\begin{figure*}
	\includegraphics[width=\textwidth]{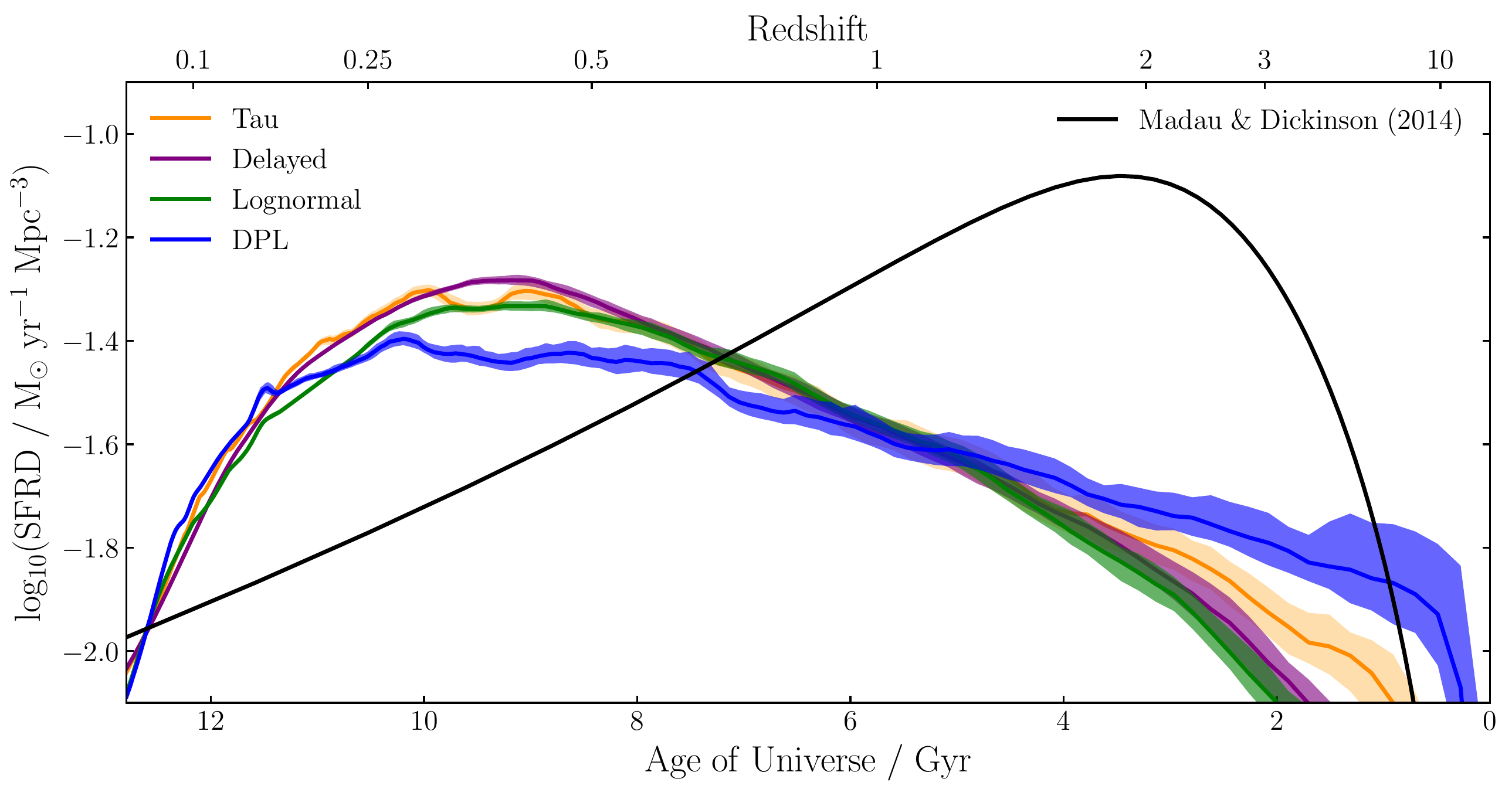}
    \caption{Redshift evolution of the cosmic SFRD as derived from the GAMA sample under the assumption of different parametric SFH models. The solid lines show the posterior medians and the shaded regions show the 16\textsuperscript{th}--84\textsuperscript{th} percentiles. The result obtained by \protect \cite{Madau2014} by measuring the SFRs of galaxies across cosmic time is shown in black. The shape of the SFRD curve can be seen to be poorly reproduced, with mass assembly occurring later in cosmic history.}\label{fig:sfrd}
\end{figure*}

\subsection{Inferred mass-weighted formation times: the shape of the inferred SFRD evolution}\label{subsect:madau_shape}

We finally consider the mass-weighted formation times we infer for our sample, by considering the implied redshift evolution of the cosmic SFRD. Our results are shown in Fig. \ref{fig:sfrd}, along with the \cite{Madau2014} result. Whilst we have demonstrated in Sections \ref{subsect:madau_smd} and \ref{subsect:madau_sfrs} that the stellar masses and SFRs we infer for our sample at $z\sim0.05$ are broadly consistent with \cite{Madau2014}, it can be seen that the cosmic SFRD evolution we infer is very different. Our SFRD curves peak significantly later in cosmic history, at $z\sim0.4$, when the Universe is $9-10$ Gyr old. This is in marked contrast to the \cite{Madau2014} curve, for which the SFRD peaks at $z\sim2$, when the Universe is $2-3$ Gyr old.

Measuring SFRs for individual high-redshift galaxies is a more direct measurement of the redshift evolution of the cosmic SFRD, and can be regarded, for the purposes of our discussion, as ground truth. Fig. \ref{fig:sfrd} therefore suggests that we significantly overestimate the mass-weighted formation times for galaxies in our sample, by as much as $\sim5$ Gyr on average.

A similar overestimation of mass-weighted formation times (underestimation of mass-weighted ages) has been previously observed in SED fitting analyses using both parametric and non-parametric SFHs (e.g. \citealt{Gallazzi2008}, \citealt{Wuyts2011}, \citealt{Leitner2012}). An analysis by \cite{Panter2003} and \cite{Heavens2004} extracted SFHs for almost 100,000 galaxies from SDSS spectra using the \textsc{Moped} code, which employs an 11-parameter non-parametric SFH model. They found that the cosmic SFRD peaks at $z\sim0.5$, when the Universe was $8-9$ Gyr old, similar to the results obtained with our parametric SFH models. However, an updated analysis by \cite{Panter2007} found that star-formation peaked in their highest-redshift bin at $z > 1.87$, demonstrating the significant impact of SED-modelling assumptions on mass-weighted formation time inferences.

In Fig. \ref{fig:fiducial_priors}, we showed that the priors imposed by our parametric SFH models favour later galaxy formation than \cite{Madau2014}. In fact, the prior medians shown on the centre-left panels of Fig. \ref{fig:fiducial_priors} correspond closely to the peaks in cosmic SFRD shown in Fig. \ref{fig:sfrd}. It seems likely, therefore, that the strong posterior constraints we obtain for cosmic SFRD evolution, favouring a late peak in cosmic SFRD, are a consequence of the strong priors imposed by our parametric SFH models, rather than the photometric data we fit. This conclusion is supported by the fact that the DPL model exhibits both a prior preference for older ages than the other models in Fig. \ref{fig:fiducial_priors}, and earlier mass assembly in Fig. \ref{fig:sfrd}.

A key goal for contemporary SED fitting analyses is to move beyond the acquisition of galaxy physical parameters at the redshift of observation to reliably infer mass-assembly histories at earlier times. This ability would be a valuable tool for explicitly linking galaxy populations at different epochs (e.g. \citealt{McLure2018, Pentericci2018}). Whilst it has long been known that tau models do not produce accurate mass-assembly histories (e.g. \citealt{Wuyts2011}; \citealt{Reddy2012}; \citealt{Pforr2012}; \citealt{Buat2014}), our results suggest that newer parameterisations, such as the lognormal and DPL models, do not significantly improve our ability to obtain realistic mass-assembly histories from the observed SED. Whilst it is possible to calibrate these models to obtain unbiased results in certain circumstances (e.g. \citealt{Carnall2018}), in order to obtain realistic mass-assembly histories for representative samples of galaxies, further consideration should be given to non-parametric and simulation-derived SFH models.

\section{conclusion}

In this work we have carried out an investigation of the effects of four parametric SFH models (exponentially declining, delayed exponentially declining, lognormal and double power law) on galaxy stellar masses, SFRs and mass-weighted formation times. We have considered:

\begin{itemize}
\item The priors imposed on physical parameters by the use of each parametric SFH model in Section \ref{sect:priors}.

\item The biases introduced when fitting mock high-SNR broad-band photometric data in Section \ref{sect:mocks}.

\item The consistency of SFHs inferred for a volume-complete, low-redshift sample of galaxies from GAMA with the cosmic SFRD evolution reported by \cite{Madau2014} in Section \ref{sect:madau}.

\end{itemize}

\noindent In Fig. \ref{fig:fiducial_priors} we demonstrate that each of these parametric models imposes relatively similar, strongly peaked priors on sSFR, which could act to tighten and shift the SFMS, depending on the details of the modelling assumptions used. All four SFH models also impose a prior preference for stellar-mass assembly at later times (younger stellar ages) than is observed to be the case through measuring galaxy SFRs at high redshift. Fig. \ref{fig:alt_tau_priors} demonstrates that changing the prior probability densities on model parameters can change the priors on physical parameters at least as significantly as changing the parametric SFH model adopted. In particular, a uniform prior on $1/\tau$ for the tau model is less informative on galaxy sSFRs than a uniform prior on $\tau$.

By fitting a mock catalogue of high-SNR broad-band photometry, we have shown in Fig. \ref{fig:mocks_1d_post} that inferred stellar masses, SFRs and mass-weighted formation times/ages are prior-dependent at levels of at least 0.1, 0.3 and 0.2 dex respectively. However, the dominant factor which determines how well the true values of these parameters can be recovered is the shape of the true SFH, rather than the parametric model being fitted. Our parametric models are all significantly limited in their ability to reproduce SFHs with strong, recent variations in SFR, and will consequently return strongly biased parameters when fitting galaxies with these SFHs.

Under the assumption that our mock catalogue is representative of real galaxy SFHs, we have shown that tau models produce more strongly biased physical parameter estimates than our other three models. However, in Fig. \ref{fig:evidence}, we demonstrate that high-SNR broad-band photometry cannot discriminate between prior beliefs about which parametric SFH models are most appropriate for describing real galaxy SFHs. This means that carefully considered, physically motivated priors are a necessary component of any SED fitting analysis.

Finally we have fitted a volume-complete sample of galaxies at $0.05 < z < 0.08$ with high-quality photometric data from the GAMA Survey. We demonstrate in Table \ref{table:smd} and Fig. \ref{fig:sfrs} that our \textsc{Bagpipes} stellar-mass and SFR measurements at $z \sim 0.05$ are consistent both with other SED fitting analyses from the literature and the SFRD curve of \cite{Madau2014}.

However, in Fig. \ref{fig:sfrd} we demonstrate that the mass-weighted formation times we infer are significantly overestimated (mass-weighted ages are underestimated), as the ensemble of our fitted SFHs predicts a much later peak in cosmic SFRD than \cite{Madau2014}. Our analysis suggests that the cosmic SFRD peaked at $z\sim0.4$, approximately 6 Gyr later than is directly observed. A comparison of Fig. \ref{fig:fiducial_priors} with Fig.  \ref{fig:sfrd} suggests that this result is a consequence of the poorly-motivated priors imposed by our parametric SFH models.

Our analyses demonstrate the challenges involved in using parametric SFH models as tools for understanding the history of galaxy stellar-mass assembly. Non-parametric SFH models are a promising alternative to the parametric forms discussed in this work. Such models both provide greater flexibility than parametric models, and allow prior beliefs to be incorporated in a more direct way. In a companion paper \citep{Leja2018a} we consider the advantages and disadvantages of such models by subjecting them to similar tests to those we have employed in this work.

The observation of samples of galaxies at different points in cosmic history and subsequent attempts to connect them are powerful tools for understanding galaxy evolution (e.g. \citealt{Wild2016, Belli2018}). However, despite the aim of understanding populations of galaxies, all current SED fitting analyses treat individual galaxies as statistically independent from each other. A more powerful approach would be to simultaneously model and fit whole populations of galaxies through the use of a hierarchical Bayesian model. 

This could be used, for example, to enforce continuity between galaxy populations at different redshifts by treating the redshift evolution of the cosmic SFRD as a prior distribution. The hyper-parameters of this prior could then be jointly constrained by samples of galaxies across a range of observed redshifts, using both instantaneous SFRs and SFHs. Under this scheme we would be able to self-consistently model the redshift evolution of the GSMF, SFMS and cosmic SFRD, as well as obtaining better constraints on the SFHs of individual galaxies through the use of well-motivated population priors (Bayesian shrinkage).

\acknowledgments

A.C.C. acknowledges the support of the UK Science and Technology Facilities Council, as well as the Scottish Universities Physics Alliance. J.L. is supported by an NSF Astronomy and Astrophysics Postdoctoral Fellowship under award AST-1701487. GAMA is a joint European-Australasian project based around a spectroscopic campaign using the Anglo-Australian Telescope. The GAMA input catalogue is based on data taken from the Sloan Digital Sky Survey and the UKIRT Infrared Deep Sky Survey. Complementary imaging of the GAMA regions is being obtained by a number of independent survey programmes including GALEX MIS, VST KiDS, VISTA VIKING, WISE, Herschel-ATLAS, GMRT and ASKAP providing UV to radio coverage. GAMA is funded by the STFC (UK), the ARC (Australia), the AAO, and the participating institutions. The GAMA website is http://www.gama-survey.org/. Based on observations made with ESO Telescopes at the La Silla Paranal Observatory under programme IDs 179.A-2004 and 177.A-3016.

\begin{appendix}

\section{Effects of different priors for the lognormal and double power law models}\label{lognorm_dpl_ppds}

As discussed in Section \ref{subsect:priors_z0}, the recent increase in the diversity of SFH models used in the literature precludes a side-by-side comparison of all options. Instead we advocate that authors who use new models should present the results of tests such as those performed in Section \ref{sect:priors}, in order to understand the priors which are being imposed on the physical parameters of interest. As the lognormal and DPL models we consider in this paper are relatively novel, we here further elaborate on the results of Section \ref{sect:priors} for these models. The aim is to provide a discussion which helps authors wishing to use these models to select appropriate prior probability densities for model parameters.

\subsection{The lognormal model}\label{subsect:app_lnorm}

\begin{figure*}[ht]
	\includegraphics[width=\textwidth]{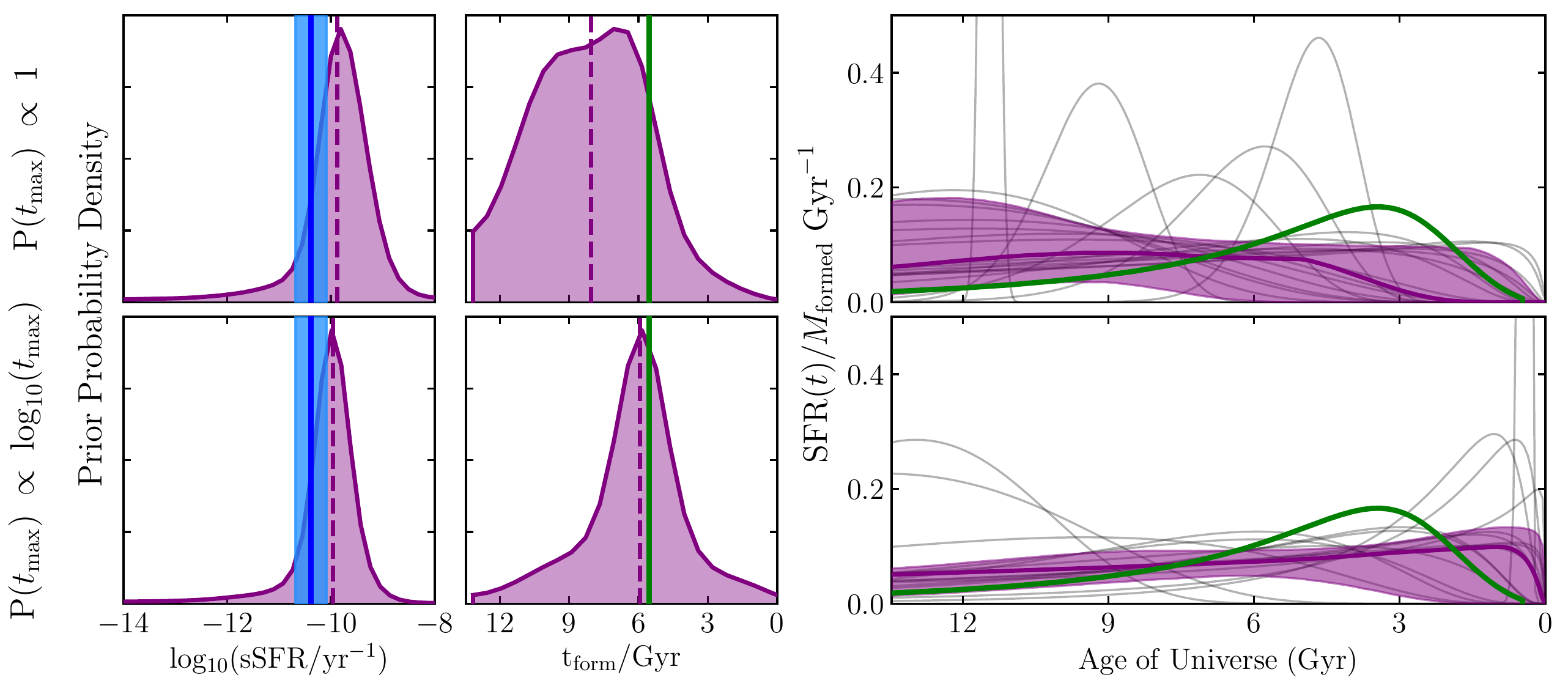}
    \caption{The effects of changing the prior on the $t_\mathrm{max}$ parameter of the lognormal SFH model. All details are as in Figure \ref{fig:fiducial_priors}.}\label{fig:extra_lognorm_priors}
\end{figure*}

A number of variations might reasonably be considered on the priors reported for this model in Table \ref{table:priors}. For example, \cite{Diemer2017} introduce a relatively complex set of priors to penalise extremely large $t_\mathrm{max}$ and $T_\mathrm{FWHM}$ values. As the uncertainties on $t_\mathrm{max}$ and $T_\mathrm{FWHM}$ span several orders of magnitude, it is reasonable to consider assigning them priors which are uniform in logarithmic, rather than linear space. This was tested in the case of $T_\mathrm{FWHM}$ and found to result in extremely narrow, bursty SFHs which, when fitted to data, typically adopted $T_\mathrm{FWHM}$ values consistent with the lower limit on the prior. As this is not thought to be a physically realistic shape for the SFHs of most galaxies, this option was discounted. Instead, in Figure \ref{fig:extra_lognorm_priors} we demonstrate the effects of imposing a prior of $\mathrm{P}(t_\mathrm{max}) \propto \mathrm{log}_{10}(t_\mathrm{max})$ between the limits quoted in Table \ref{table:priors}, as opposed to the original uniform prior shown on Fig. \ref{fig:fiducial_priors}. This can be seen to significantly change the shape prior, encoding a preference for earlier formation, which brings the prior on $t_\mathrm{form}$ into better agreement with \cite{Madau2014}. However, this change also significantly narrows the priors on SFH shape and $t_\mathrm{form}$, reducing the ability of this prior to describe the diversity of possible SFH shapes.

\subsection{The double power law model}

\begin{figure*}[ht]
	\includegraphics[width=\textwidth]{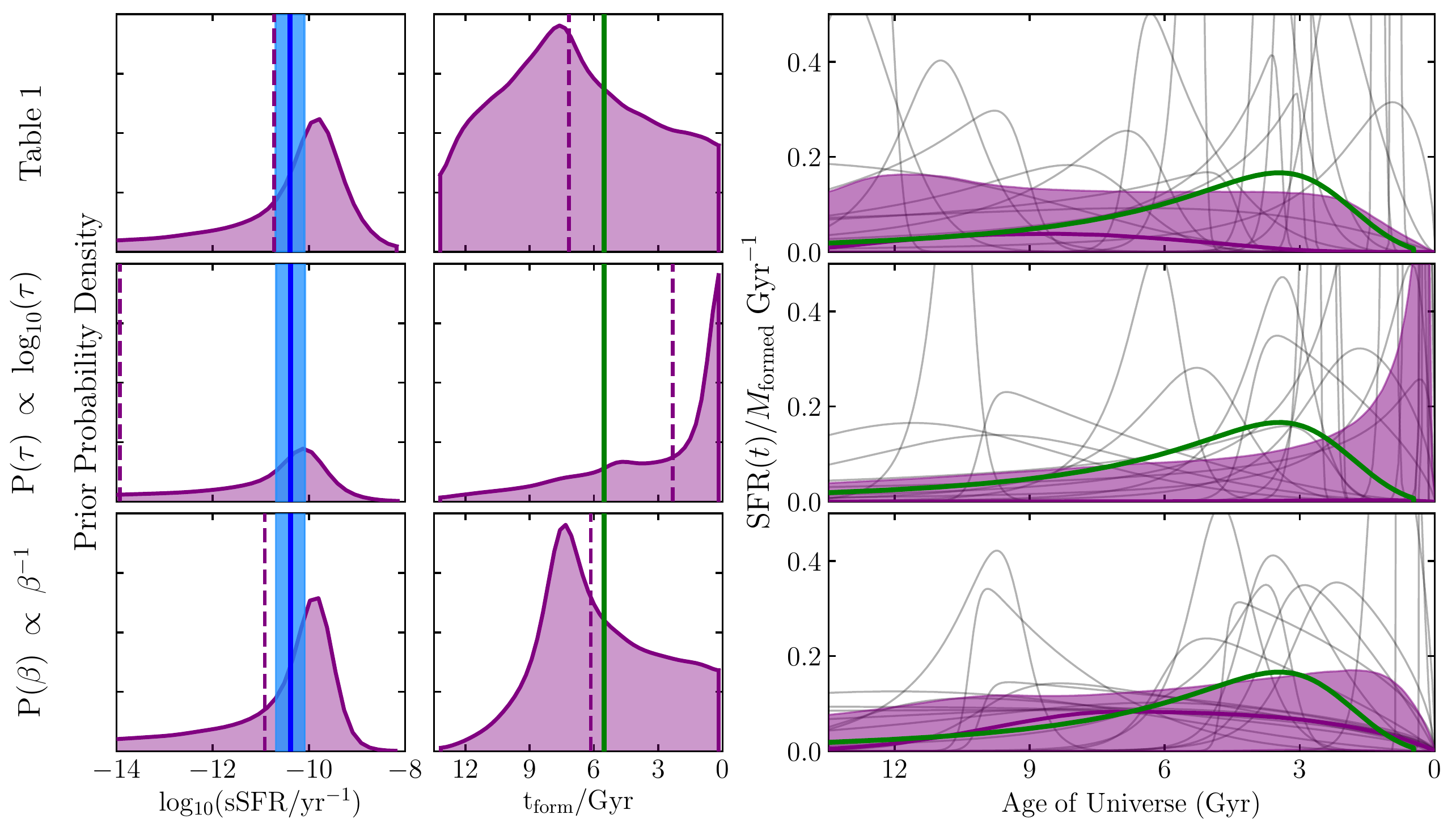}
    \caption{The effects of changing the priors on the $\tau$ and $\beta$ parameters of the DPL SFH model. All details are as in Figure \ref{fig:fiducial_priors}.}\label{fig:extra_dpl_priors}
\end{figure*}

The priors for the DPL model listed in Table \ref{table:priors} were arrived at following extensive experimentation with different options, partially detailed in \cite{Carnall2018}. The shape of the SFH changes roughly uniformly with the logarithm of $\alpha$ and $\beta$ between the limits of 0.1 and 1000. Outside of this range, further increase or decrease of these parameters results in no further change to the SFH shape, as the slopes are essentially horizontal for values $< 0.1$, or vertical for values $> 1000$. Using uniform priors on $\alpha$ and $\beta$ leads to the same effect as was discussed for the $T_\mathrm{FWHM}$ parameter of the lognormal model in Appendix \ref{subsect:app_lnorm}, with extremely bursty SFHs being assigned the bulk of the prior mass.

Given the success achieved in emphasising earlier formation observed in Fig. \ref{fig:extra_lognorm_priors}, we perform a similar test to Appendix \ref{subsect:app_lnorm} by changing the prior on the $\tau$ parameter from the initial uniform prior to $\mathrm{P}(\tau) \propto \mathrm{log}_{10}(\tau)$. The results of this test are shown in the middle row of Fig. \ref{fig:extra_dpl_priors}. It can be seen that this change overweights models which form the majority of their stars at very early times, producing a prior preference for older stellar populations than the \cite{Madau2014} curve. We therefore suggest retaining a uniform prior on $\tau$ for the DPL model.

The main advantage of the DPL model is the ability to decouple the early and late time SFHs, with separate parameters controlling the slopes of each. With this in mind, and considering the results of Fig. \ref{fig:fiducial_priors}, we assess the possibility of changing the prior on the rising slope from $\mathrm{P}(\beta) \propto \mathrm{log}_{10}(\beta)$ to $\mathrm{P}(\beta) \propto 1/\beta$. This change lends more prior weight to flatter rising slopes, thus causing star-formation to be more extended back towards earlier times before it peaks. These results are shown in the bottom row of Fig. \ref{fig:extra_dpl_priors}. This is probably the most promising of the alternatives we have considered to the priors listed in Table \ref{table:priors}, with the average of the $t_\mathrm{form}$ prior in good agreement with \cite{Madau2014}. However the average SFH shape can still be seen to be quite different from \cite{Madau2014}, with the sSFR prior also retaining a fairly strong peak.

As discussed in Section \ref{subsect:mocks_information} and demonstrated in this Appendix, the process of tuning a parametric SFH model involves a large amount of trial and error, with no clear physical link between the priors chosen and the physics the SFH model represents. We therefore suggest that non-parametric SFH models are a better choice, as they allow physical information to be included in the prior in a more direct way. For further information and examples see \cite{Leja2018a}.

\end{appendix}

\bibliographystyle{aasjournal}
\bibliography{carnall2018} 

\begin{thebibliography}{}
\expandafter\ifx\csname natexlab\endcsname\relax\def\natexlab#1{#1}\fi
\providecommand{\url}[1]{\href{#1}{#1}}
\providecommand{\dodoi}[1]{doi:~\href{http://doi.org/#1}{\nolinkurl{#1}}}
\providecommand{\doeprint}[1]{\href{http://ascl.net/#1}{\nolinkurl{http://ascl.net/#1}}}
\providecommand{\doarXiv}[1]{\href{https://arxiv.org/abs/#1}{\nolinkurl{https://arxiv.org/abs/#1}}}

\bibitem[{{Acquaviva} {et~al.}(2011){Acquaviva}, {Gawiser}, \&
  {Guaita}}]{Acquaviva2011}
{Acquaviva}, V., {Gawiser}, E., \& {Guaita}, L. 2011, \apj, 737, 47,
  \dodoi{10.1088/0004-637X/737/2/47}

\bibitem[{{Baldry} {et~al.}(2018){Baldry}, {Liske}, {Brown}, {Robotham},
  {Driver}, {Dunne}, {Alpaslan}, {Brough}, {Cluver}, {Eardley}, {Farrow},
  {Heymans}, {Hildebrandt}, {Hopkins}, {Kelvin}, {Loveday}, {Moffett},
  {Norberg}, {Owers}, {Taylor}, {Wright}, {Bamford}, {Bland-Hawthorn},
  {Bourne}, {Bremer}, {Colless}, {Conselice}, {Croom}, {Davies}, {Foster},
  {Grootes}, {Holwerda}, {Jones}, {Kafle}, {Kuijken}, {Lara-Lopez},
  {L{\'o}pez-S{\'a}nchez}, {Meyer}, {Phillipps}, {Sutherland}, {van Kampen}, \&
  {Wilkins}}]{Baldry2018}
{Baldry}, I.~K., {Liske}, J., {Brown}, M.~J.~I., {et~al.} 2018, \mnras, 474,
  3875, \dodoi{10.1093/mnras/stx3042}

\bibitem[{{Behroozi} {et~al.}(2013){Behroozi}, {Wechsler}, \&
  {Conroy}}]{Behroozi2013}
{Behroozi}, P.~S., {Wechsler}, R.~H., \& {Conroy}, C. 2013, \apj, 770, 57,
  \dodoi{10.1088/0004-637X/770/1/57}

\bibitem[{{Belli} {et~al.}(2018){Belli}, {Newman}, \& {Ellis}}]{Belli2018}
{Belli}, S., {Newman}, A.~B., \& {Ellis}, R.~S. 2018, ArXiv e-prints.
\newblock \doarXiv{1810.00008}

\bibitem[{{Brammer} {et~al.}(2008){Brammer}, {van Dokkum}, \&
  {Coppi}}]{Brammer2008}
{Brammer}, G.~B., {van Dokkum}, P.~G., \& {Coppi}, P. 2008, \apj, 686, 1503,
  \dodoi{10.1086/591786}

\bibitem[{{Buat} {et~al.}(2014){Buat}, {Heinis}, {Boquien}, {Burgarella},
  {Charmandaris}, {Boissier}, {Boselli}, {Le Borgne}, \& {Morrison}}]{Buat2014}
{Buat}, V., {Heinis}, S., {Boquien}, M., {et~al.} 2014, \aap, 561, A39,
  \dodoi{10.1051/0004-6361/201322081}

\bibitem[{{Calzetti} {et~al.}(2000){Calzetti}, {Armus}, {Bohlin}, {Kinney},
  {Koornneef}, \& {Storchi-Bergmann}}]{Calzetti2000}
{Calzetti}, D., {Armus}, L., {Bohlin}, R.~C., {et~al.} 2000, \apj, 533, 682,
  \dodoi{10.1086/308692}

\bibitem[{{Cappellari}(2017)}]{Cappellari2017}
{Cappellari}, M. 2017, MNRAS, 466, 798, \dodoi{10.1093/mnras/stw3020}

\bibitem[{{Carnall} {et~al.}(2018){Carnall}, {McLure}, {Dunlop}, \&
  {Dav{\'e}}}]{Carnall2018}
{Carnall}, A.~C., {McLure}, R.~J., {Dunlop}, J.~S., \& {Dav{\'e}}, R. 2018,
  \mnras, \dodoi{10.1093/mnras/sty2169}

\bibitem[{{Chauke} {et~al.}(2018){Chauke}, {van der Wel}, {Pacifici},
  {Bezanson}, {Wu}, {Gallazzi}, {Noeske}, {Straatman}, {Mu{\~n}os-Mateos},
  {Franx}, {Bari{\v s}i{\'c}}, {Bell}, {Brammer}, {Calhau}, {van Houdt},
  {Labb{\'e}}, {Maseda}, {Muzzin}, {Rix}, \& {Sobral}}]{Chauke2018}
{Chauke}, P., {van der Wel}, A., {Pacifici}, C., {et~al.} 2018, \apj, 861, 13,
  \dodoi{10.3847/1538-4357/aac324}

\bibitem[{{Chevallard} \& {Charlot}(2016)}]{Chevallard2016}
{Chevallard}, J., \& {Charlot}, S. 2016, \mnras, 462, 1415,
  \dodoi{10.1093/mnras/stw1756}

\bibitem[{{Chevallard} {et~al.}(2017){Chevallard}, {Curtis-Lake}, {Charlot},
  {Ferruit}, {Giardino}, {Franx}, {Maseda}, {Amorin}, {Arribas}, {Bunker},
  {Carniani}, {Husemann}, {Jakobsen}, {Maiolino}, {Pforr}, {Rawle}, {Rix},
  {Smit}, \& {Willott}}]{Chevallard2017}
{Chevallard}, J., {Curtis-Lake}, E., {Charlot}, S., {et~al.} 2017, ArXiv
  e-prints.
\newblock \doarXiv{1711.07481}

\bibitem[{{Cid Fernandes} {et~al.}(2005){Cid Fernandes}, {Mateus}, {Sodr{\'e}},
  {Stasi{\'n}ska}, \& {Gomes}}]{CidFernandes2005}
{Cid Fernandes}, R., {Mateus}, A., {Sodr{\'e}}, L., {Stasi{\'n}ska}, G., \&
  {Gomes}, J.~M. 2005, \mnras, 358, 363,
  \dodoi{10.1111/j.1365-2966.2005.08752.x}

\bibitem[{{Ciesla} {et~al.}(2017){Ciesla}, {Elbaz}, \& {Fensch}}]{Ciesla2017}
{Ciesla}, L., {Elbaz}, D., \& {Fensch}, J. 2017, \aap, 608, A41,
  \dodoi{10.1051/0004-6361/201731036}

\bibitem[{{Ciesla} {et~al.}(2015){Ciesla}, {Charmandaris}, {Georgakakis},
  {Bernhard}, {Mitchell}, {Buat}, {Elbaz}, {LeFloc'h}, {Lacey}, {Magdis}, \&
  {Xilouris}}]{Ciesla2015}
{Ciesla}, L., {Charmandaris}, V., {Georgakakis}, A., {et~al.} 2015, \aap, 576,
  A10, \dodoi{10.1051/0004-6361/201425252}

\bibitem[{{Cohn}(2018)}]{Cohn2018}
{Cohn}, J.~D. 2018, \mnras, 478, 2291, \dodoi{10.1093/mnras/sty1148}

\bibitem[{{Conroy}(2013)}]{Conroy2013}
{Conroy}, C. 2013, \araa, 51, 393, \dodoi{10.1146/annurev-astro-082812-141017}

\bibitem[{{da Cunha} {et~al.}(2008){da Cunha}, {Charlot}, \&
  {Elbaz}}]{daCunha2008}
{da Cunha}, E., {Charlot}, S., \& {Elbaz}, D. 2008, \mnras, 388, 1595,
  \dodoi{10.1111/j.1365-2966.2008.13535.x}

\bibitem[{{Dav{\'e}} {et~al.}(2016){Dav{\'e}}, {Thompson}, \&
  {Hopkins}}]{Dave2016}
{Dav{\'e}}, R., {Thompson}, R., \& {Hopkins}, P.~F. 2016, \mnras, 462, 3265,
  \dodoi{10.1093/mnras/stw1862}

\bibitem[{{Diemer} {et~al.}(2017){Diemer}, {Sparre}, {Abramson}, \&
  {Torrey}}]{Diemer2017}
{Diemer}, B., {Sparre}, M., {Abramson}, L.~E., \& {Torrey}, P. 2017, \apj, 839,
  26, \dodoi{10.3847/1538-4357/aa68e5}

\bibitem[{{Dom{\'{\i}}nguez} {et~al.}(2013){Dom{\'{\i}}nguez}, {Siana},
  {Henry}, {Scarlata}, {Bedregal}, {Malkan}, {Atek}, {Ross}, {Colbert},
  {Teplitz}, {Rafelski}, {McCarthy}, {Bunker}, {Hathi}, {Dressler}, {Martin},
  \& {Masters}}]{Dominguez2013}
{Dom{\'{\i}}nguez}, A., {Siana}, B., {Henry}, A.~L., {et~al.} 2013, \apj, 763,
  145, \dodoi{10.1088/0004-637X/763/2/145}

\bibitem[{{Draine} \& {Li}(2007)}]{Draine2007}
{Draine}, B.~T., \& {Li}, A. 2007, \apj, 657, 810, \dodoi{10.1086/511055}

\bibitem[{{Driver} {et~al.}(2009){Driver}, {Norberg}, {Baldry}, {Bamford},
  {Hopkins}, {Liske}, {Loveday}, {Peacock}, {Hill}, {Kelvin}, {Robotham},
  {Cross}, {Parkinson}, {Prescott}, {Conselice}, {Dunne}, {Brough}, {Jones},
  {Sharp}, {van Kampen}, {Oliver}, {Roseboom}, {Bland-Hawthorn}, {Croom},
  {Ellis}, {Cameron}, {Cole}, {Frenk}, {Couch}, {Graham}, {Proctor}, {De
  Propris}, {Doyle}, {Edmondson}, {Nichol}, {Thomas}, {Eales}, {Jarvis},
  {Kuijken}, {Lahav}, {Madore}, {Seibert}, {Meyer}, {Staveley-Smith},
  {Phillipps}, {Popescu}, {Sansom}, {Sutherland}, {Tuffs}, \&
  {Warren}}]{Driver2009}
{Driver}, S.~P., {Norberg}, P., {Baldry}, I.~K., {et~al.} 2009, Astronomy and
  Geophysics, 50, 5.12, \dodoi{10.1111/j.1468-4004.2009.50512.x}

\bibitem[{{Driver} {et~al.}(2016){Driver}, {Wright}, {Andrews}, {Davies},
  {Kafle}, {Lange}, {Moffett}, {Mannering}, {Robotham}, {Vinsen}, {Alpaslan},
  {Andrae}, {Baldry}, {Bauer}, {Bamford}, {Bland-Hawthorn}, {Bourne}, {Brough},
  {Brown}, {Cluver}, {Croom}, {Colless}, {Conselice}, {da Cunha}, {De Propris},
  {Drinkwater}, {Dunne}, {Eales}, {Edge}, {Frenk}, {Graham}, {Grootes},
  {Holwerda}, {Hopkins}, {Ibar}, {van Kampen}, {Kelvin}, {Jarrett}, {Jones},
  {Lara-Lopez}, {Liske}, {Lopez-Sanchez}, {Loveday}, {Maddox}, {Madore},
  {Mahajan}, {Meyer}, {Norberg}, {Penny}, {Phillipps}, {Popescu}, {Tuffs},
  {Peacock}, {Pimbblet}, {Prescott}, {Rowlands}, {Sansom}, {Seibert}, {Smith},
  {Sutherland}, {Taylor}, {Valiante}, {Vazquez-Mata}, {Wang}, {Wilkins}, \&
  {Williams}}]{Driver2016}
{Driver}, S.~P., {Wright}, A.~H., {Andrews}, S.~K., {et~al.} 2016, \mnras, 455,
  3911, \dodoi{10.1093/mnras/stv2505}

\bibitem[{{Driver} {et~al.}(2018){Driver}, {Andrews}, {da Cunha}, {Davies},
  {Lagos}, {Robotham}, {Vinsen}, {Wright}, {Alpaslan}, {Bland-Hawthorn},
  {Bourne}, {Brough}, {Bremer}, {Cluver}, {Colless}, {Conselice}, {Dunne},
  {Eales}, {Gomez}, {Holwerda}, {Hopkins}, {Kafle}, {Kelvin}, {Loveday},
  {Liske}, {Maddox}, {Phillipps}, {Pimbblet}, {Rowlands}, {Sansom}, {Taylor},
  {Wang}, \& {Wilkins}}]{Driver2018}
{Driver}, S.~P., {Andrews}, S.~K., {da Cunha}, E., {et~al.} 2018, \mnras, 475,
  2891, \dodoi{10.1093/mnras/stx2728}

\bibitem[{{Feroz} {et~al.}(2009){Feroz}, {Hobson}, \& {Bridges}}]{Feroz2009}
{Feroz}, F., {Hobson}, M.~P., \& {Bridges}, M. 2009, \mnras, 398, 1601,
  \dodoi{10.1111/j.1365-2966.2009.14548.x}

\bibitem[{{Feroz} {et~al.}(2013){Feroz}, {Hobson}, {Cameron}, \&
  {Pettitt}}]{Feroz2013}
{Feroz}, F., {Hobson}, M.~P., {Cameron}, E., \& {Pettitt}, A.~N. 2013, ArXiv
  e-prints.
\newblock \doarXiv{1306.2144}

\bibitem[{{Foreman-Mackey} {et~al.}(2013){Foreman-Mackey}, {Hogg}, {Lang}, \&
  {Goodman}}]{Foreman-Mackey2013}
{Foreman-Mackey}, D., {Hogg}, D.~W., {Lang}, D., \& {Goodman}, J. 2013, \pasp,
  125, 306, \dodoi{10.1086/670067}

\bibitem[{{Gallazzi} {et~al.}(2008){Gallazzi}, {Brinchmann}, {Charlot}, \&
  {White}}]{Gallazzi2008}
{Gallazzi}, A., {Brinchmann}, J., {Charlot}, S., \& {White}, S.~D.~M. 2008,
  \mnras, 383, 1439, \dodoi{10.1111/j.1365-2966.2007.12632.x}

\bibitem[{{Gallazzi} {et~al.}(2005){Gallazzi}, {Charlot}, {Brinchmann},
  {White}, \& {Tremonti}}]{Gallazzi2005}
{Gallazzi}, A., {Charlot}, S., {Brinchmann}, J., {White}, S.~D.~M., \&
  {Tremonti}, C.~A. 2005, \mnras, 362, 41,
  \dodoi{10.1111/j.1365-2966.2005.09321.x}

\bibitem[{{Garn} \& {Best}(2010)}]{Garn2010}
{Garn}, T., \& {Best}, P.~N. 2010, \mnras, 409, 421,
  \dodoi{10.1111/j.1365-2966.2010.17321.x}

\bibitem[{{Gladders} {et~al.}(2013){Gladders}, {Oemler}, {Dressler},
  {Poggianti}, {Vulcani}, \& {Abramson}}]{Gladders2013}
{Gladders}, M.~D., {Oemler}, A., {Dressler}, A., {et~al.} 2013, \apj, 770, 64,
  \dodoi{10.1088/0004-637X/770/1/64}

\bibitem[{{Glazebrook} {et~al.}(2017){Glazebrook}, {Schreiber}, {Labb{\'e}},
  {Nanayakkara}, {Kacprzak}, {Oesch}, {Papovich}, {Spitler}, {Straatman},
  {Tran}, \& {Yuan}}]{Glazebrook2017}
{Glazebrook}, K., {Schreiber}, C., {Labb{\'e}}, I., {et~al.} 2017, \nat, 544,
  71, \dodoi{10.1038/nature21680}

\bibitem[{{Goodman} \& {Weare}(2010)}]{Goodman2010}
{Goodman}, J., \& {Weare}, J. 2010, Communications in Applied Mathematics and
  Computational Science, Vol.~5, No.~1, p.~65-80, 2010, 5, 65,
  \dodoi{10.2140/camcos.2010.5.65}

\bibitem[{{Heavens} {et~al.}(2004){Heavens}, {Panter}, {Jimenez}, \&
  {Dunlop}}]{Heavens2004}
{Heavens}, A., {Panter}, B., {Jimenez}, R., \& {Dunlop}, J. 2004, \nat, 428,
  625, \dodoi{10.1038/nature02474}

\bibitem[{{Iyer} \& {Gawiser}(2017)}]{Iyer2017}
{Iyer}, K., \& {Gawiser}, E. 2017, \apj, 838, 127,
  \dodoi{10.3847/1538-4357/aa63f0}

\bibitem[{{Kennicutt} \& {Evans}(2012)}]{Kennicutt2012}
{Kennicutt}, R.~C., \& {Evans}, N.~J. 2012, \araa, 50, 531,
  \dodoi{10.1146/annurev-astro-081811-125610}

\bibitem[{{Kroupa} \& {Boily}(2002)}]{Kroupa2002}
{Kroupa}, P., \& {Boily}, C.~M. 2002, \mnras, 336, 1188,
  \dodoi{10.1046/j.1365-8711.2002.05848.x}

\bibitem[{{Lange} {et~al.}(2015){Lange}, {Driver}, {Robotham}, {Kelvin},
  {Graham}, {Alpaslan}, {Andrews}, {Baldry}, {Bamford}, {Bland-Hawthorn},
  {Brough}, {Cluver}, {Conselice}, {Davies}, {Haeussler}, {Konstantopoulos},
  {Loveday}, {Moffett}, {Norberg}, {Phillipps}, {Taylor},
  {L{\'o}pez-S{\'a}nchez}, \& {Wilkins}}]{Lange2015}
{Lange}, R., {Driver}, S.~P., {Robotham}, A.~S.~G., {et~al.} 2015, \mnras, 447,
  2603, \dodoi{10.1093/mnras/stu2467}

\bibitem[{{Lee} {et~al.}(2009){Lee}, {Idzi}, {Ferguson}, {Somerville},
  {Wiklind}, \& {Giavalisco}}]{Lee2009}
{Lee}, S.-K., {Idzi}, R., {Ferguson}, H.~C., {et~al.} 2009, \apjs, 184, 100,
  \dodoi{10.1088/0067-0049/184/1/100}

\bibitem[{{Leitner}(2012)}]{Leitner2012}
{Leitner}, S.~N. 2012, \apj, 745, 149, \dodoi{10.1088/0004-637X/745/2/149}

\bibitem[{{Leja} {et~al.}(2018){Leja}, {Carnall}, {Johnson}, {Conroy}, \&
  {Speagle}}]{Leja2018a}
{Leja}, J., {Carnall}, A.~C., {Johnson}, B.~D., {Conroy}, C., \& {Speagle},
  J.~S. 2018, arXiv e-prints.
\newblock \doarXiv{1811.03637}

\bibitem[{{Leja} {et~al.}(2017){Leja}, {Johnson}, {Conroy}, {van Dokkum}, \&
  {Byler}}]{Leja2017}
{Leja}, J., {Johnson}, B.~D., {Conroy}, C., {van Dokkum}, P.~G., \& {Byler}, N.
  2017, \apj, 837, 170, \dodoi{10.3847/1538-4357/aa5ffe}

\bibitem[{{Leja} {et~al.}(2015){Leja}, {van Dokkum}, {Franx}, \&
  {Whitaker}}]{Leja2015}
{Leja}, J., {van Dokkum}, P.~G., {Franx}, M., \& {Whitaker}, K.~E. 2015, \apj,
  798, 115, \dodoi{10.1088/0004-637X/798/2/115}

\bibitem[{{Madau} \& {Dickinson}(2014)}]{Madau2014}
{Madau}, P., \& {Dickinson}, M. 2014, \araa, 52, 415,
  \dodoi{10.1146/annurev-astro-081811-125615}

\bibitem[{{McLure} {et~al.}(2011){McLure}, {Dunlop}, {de Ravel}, {Cirasuolo},
  {Ellis}, {Schenker}, {Robertson}, {Koekemoer}, {Stark}, \&
  {Bowler}}]{McLure2011}
{McLure}, R.~J., {Dunlop}, J.~S., {de Ravel}, L., {et~al.} 2011, \mnras, 418,
  2074, \dodoi{10.1111/j.1365-2966.2011.19626.x}

\bibitem[{{McLure} {et~al.}(2018){McLure}, {Pentericci}, {Cimatti}, {Dunlop},
  {Elbaz}, {Fontana}, {Nandra}, {Amorin}, {Bolzonella}, {Bongiorno}, {Carnall},
  {Castellano}, {Cirasuolo}, {Cucciati}, {Cullen}, {De Barros}, {Finkelstein},
  {Fontanot}, {Franzetti}, {Fumana}, {Gargiulo}, {Garilli}, {Guaita},
  {Hartley}, {Iovino}, {Jarvis}, {Juneau}, {Karman}, {Maccagni}, {Marchi},
  {M{\'a}rmol-Queralt{\'o}}, {Pompei}, {Pozzetti}, {Scodeggio}, {Sommariva},
  {Talia}, {Almaini}, {Balestra}, {Bardelli}, {Bell}, {Bourne}, {Bowler},
  {Brusa}, {Buitrago}, {Caputi}, {Cassata}, {Charlot}, {Citro}, {Cresci},
  {Cristiani}, {Curtis-Lake}, {Dickinson}, {Fazio}, {Ferguson}, {Fiore},
  {Franco}, {Fynbo}, {Galametz}, {Georgakakis}, {Giavalisco}, {Grazian},
  {Hathi}, {Jung}, {Kim}, {Koekemoer}, {Khusanova}, {Le F{\`e}vre}, {Lotz},
  {Mannucci}, {Maltby}, {Matsuoka}, {McLeod}, {Mendez-Hernandez},
  {Mendez-Abreu}, {Mignoli}, {Moresco}, {Mortlock}, {Nonino}, {Pannella},
  {Papovich}, {Popesso}, {Rosario}, {Salvato}, {Santini}, {Schaerer},
  {Schreiber}, {Stark}, {Tasca}, {Thomas}, {Treu}, {Vanzella}, {Wild},
  {Williams}, {Zamorani}, \& {Zucca}}]{McLure2018}
{McLure}, R.~J., {Pentericci}, L., {Cimatti}, A., {et~al.} 2018, \mnras, 479,
  25, \dodoi{10.1093/mnras/sty1213}

\bibitem[{{Merlin} {et~al.}(2018){Merlin}, {Fontana}, {Castellano}, {Santini},
  {Torelli}, {Boutsia}, {Wang}, {Grazian}, {Pentericci}, {Schreiber}, {Ciesla},
  {McLure}, {Derriere}, {Dunlop}, \& {Elbaz}}]{Merlin2018}
{Merlin}, E., {Fontana}, A., {Castellano}, M., {et~al.} 2018, \mnras, 473,
  2098, \dodoi{10.1093/mnras/stx2385}

\bibitem[{{Mobasher} {et~al.}(2015){Mobasher}, {Dahlen}, {Ferguson},
  {Acquaviva}, {Barro}, {Finkelstein}, {Fontana}, {Gruetzbauch}, {Johnson},
  {Lu}, {Papovich}, {Pforr}, {Salvato}, {Somerville}, {Wiklind}, {Wuyts},
  {Ashby}, {Bell}, {Conselice}, {Dickinson}, {Faber}, {Fazio}, {Finlator},
  {Galametz}, {Gawiser}, {Giavalisco}, {Grazian}, {Grogin}, {Guo}, {Hathi},
  {Kocevski}, {Koekemoer}, {Koo}, {Newman}, {Reddy}, {Santini}, \&
  {Wechsler}}]{Mobasher2015}
{Mobasher}, B., {Dahlen}, T., {Ferguson}, H.~C., {et~al.} 2015, \apj, 808, 101,
  \dodoi{10.1088/0004-637X/808/1/101}

\bibitem[{{Mortlock} {et~al.}(2017){Mortlock}, {McLure}, {Bowler}, {McLeod},
  {M{\'a}rmol-Queralt{\'o}}, {Parsa}, {Dunlop}, \& {Bruce}}]{Mortlock2017}
{Mortlock}, A., {McLure}, R.~J., {Bowler}, R. A.~A., {et~al.} 2017, \mnras,
  465, 672, \dodoi{10.1093/mnras/stw2728}

\bibitem[{{Nelson} {et~al.}(2018){Nelson}, {Pillepich}, {Springel},
  {Weinberger}, {Hernquist}, {Pakmor}, {Genel}, {Torrey}, {Vogelsberger},
  {Kauffmann}, {Marinacci}, \& {Naiman}}]{Nelson2018}
{Nelson}, D., {Pillepich}, A., {Springel}, V., {et~al.} 2018, \mnras, 475, 624,
  \dodoi{10.1093/mnras/stx3040}

\bibitem[{{Ocvirk} {et~al.}(2006){Ocvirk}, {Pichon}, {Lan{\c c}on}, \&
  {Thi{\'e}baut}}]{Ocvirk2006}
{Ocvirk}, P., {Pichon}, C., {Lan{\c c}on}, A., \& {Thi{\'e}baut}, E. 2006,
  \mnras, 365, 46, \dodoi{10.1111/j.1365-2966.2005.09182.x}

\bibitem[{{Pacifici} {et~al.}(2012){Pacifici}, {Charlot}, {Blaizot}, \&
  {Brinchmann}}]{Pacifici2012}
{Pacifici}, C., {Charlot}, S., {Blaizot}, J., \& {Brinchmann}, J. 2012, \mnras,
  421, 2002, \dodoi{10.1111/j.1365-2966.2012.20431.x}

\bibitem[{{Pacifici} {et~al.}(2015){Pacifici}, {da Cunha}, {Charlot}, {Rix},
  {Fumagalli}, {Wel}, {Franx}, {Maseda}, {van Dokkum}, {Brammer}, {Momcheva},
  {Skelton}, {Whitaker}, {Leja}, {Lundgren}, {Kassin}, \& {Yi}}]{Pacifici2015}
{Pacifici}, C., {da Cunha}, E., {Charlot}, S., {et~al.} 2015, \mnras, 447, 786,
  \dodoi{10.1093/mnras/stu2447}

\bibitem[{{Pacifici} {et~al.}(2016){Pacifici}, {Kassin}, {Weiner}, {Holden},
  {Gardner}, {Faber}, {Ferguson}, {Koo}, {Primack}, {Bell}, {Dekel}, {Gawiser},
  {Giavalisco}, {Rafelski}, {Simons}, {Barro}, {Croton}, {Dav{\'e}}, {Fontana},
  {Grogin}, {Koekemoer}, {Lee}, {Salmon}, {Somerville}, \&
  {Behroozi}}]{Pacifici2016}
{Pacifici}, C., {Kassin}, S.~A., {Weiner}, B.~J., {et~al.} 2016, \apj, 832, 79,
  \dodoi{10.3847/0004-637X/832/1/79}

\bibitem[{{Panter} {et~al.}(2003){Panter}, {Heavens}, \&
  {Jimenez}}]{Panter2003}
{Panter}, B., {Heavens}, A.~F., \& {Jimenez}, R. 2003, \mnras, 343, 1145,
  \dodoi{10.1046/j.1365-8711.2003.06722.x}

\bibitem[{{Panter} {et~al.}(2007){Panter}, {Jimenez}, {Heavens}, \&
  {Charlot}}]{Panter2007}
{Panter}, B., {Jimenez}, R., {Heavens}, A.~F., \& {Charlot}, S. 2007, \mnras,
  378, 1550, \dodoi{10.1111/j.1365-2966.2007.11909.x}

\bibitem[{{Pentericci} {et~al.}(2018){Pentericci}, {Garilli}, {Cucciati},
  {Franzetti}, {Iovino}, {Amorin}, {Bolzonella}, {Bongiorno}, {Carnall},
  {Castellano}, {Cimatti}, {Cirasuolo}, {Cullen}, {DeBarros}, {Dunlop},
  {Elbaz}, {Finkelstein}, {Fontana}, {Fontanot}, {Fumana}, {Gargiulo},
  {Guaita}, {Hartley}, {Jarvis}, {Juneau}, {Karman}, {Maccagni}, {Marchi},
  {Marmol-Queralto}, {Nandra}, {Pompei}, {Pozzetti}, {Scodeggio}, {Sommariva},
  {Talia}, {Almaini}, {Balestra}, {Bardelli}, {Bell}, {Bourne}, {Bowler},
  {Brusa}, {Buitrago}, {Caputi}, {Cassata}, {Charlot}, {Citro}, {Cresci},
  {Cristiani}, {Curtis-Lake}, {Dickinson}, {Faber}, {Fazio}, {Ferguson},
  {Fiore}, {Franco}, {Fynbo}, {Galametz}, {Georgakakis}, {Giavalisco},
  {Grazian}, {Hathi}, {Jung}, {Kim}, {Koekemoer}, {Khusanova}, {Le F{\`e}vre},
  {Lotz}, {Mannucci}, {Maltby}, {Matsuoka}, {McLeod}, {Mendez-Hernandez},
  {Mendez-Abreu}, {Mignoli}, {Moresco}, {Mortlock}, {Nonino}, {Pannella},
  {Papovich}, {Popesso}, {Rosario}, {Rosati}, {Salvato}, {Santini}, {Schaerer},
  {Schreiber}, {Stark}, {Tasca}, {Thomas}, {Treu}, {Vanzella}, {Wild},
  {Williams}, {Zamorani}, \& {Zucca}}]{Pentericci2018}
{Pentericci}, L., {Garilli}, R.~J.~M.~B., {Cucciati}, O., {et~al.} 2018, ArXiv
  e-prints.
\newblock \doarXiv{1803.07373}

\bibitem[{{Pforr} {et~al.}(2012){Pforr}, {Maraston}, \& {Tonini}}]{Pforr2012}
{Pforr}, J., {Maraston}, C., \& {Tonini}, C. 2012, \mnras, 422, 3285,
  \dodoi{10.1111/j.1365-2966.2012.20848.x}

\bibitem[{{Reddy} {et~al.}(2012){Reddy}, {Pettini}, {Steidel}, {Shapley},
  {Erb}, \& {Law}}]{Reddy2012}
{Reddy}, N.~A., {Pettini}, M., {Steidel}, C.~C., {et~al.} 2012, \apj, 754, 25,
  \dodoi{10.1088/0004-637X/754/1/25}

\bibitem[{{Salim} {et~al.}(2016){Salim}, {Lee}, {Janowiecki}, {da Cunha},
  {Dickinson}, {Boquien}, {Burgarella}, {Salzer}, \& {Charlot}}]{Salim2016}
{Salim}, S., {Lee}, J.~C., {Janowiecki}, S., {et~al.} 2016, \apjs, 227, 2,
  \dodoi{10.3847/0067-0049/227/1/2}

\bibitem[{{Salmon} {et~al.}(2015){Salmon}, {Papovich}, {Finkelstein}, {Tilvi},
  {Finlator}, {Behroozi}, {Dahlen}, {Dav{\'e}}, {Dekel}, {Dickinson},
  {Ferguson}, {Giavalisco}, {Long}, {Lu}, {Mobasher}, {Reddy}, {Somerville}, \&
  {Wechsler}}]{Salmon2015}
{Salmon}, B., {Papovich}, C., {Finkelstein}, S.~L., {et~al.} 2015, \apj, 799,
  183, \dodoi{10.1088/0004-637X/799/2/183}

\bibitem[{{Salmon} {et~al.}(2016){Salmon}, {Papovich}, {Long}, {Willner},
  {Finkelstein}, {Ferguson}, {Dickinson}, {Duncan}, {Faber}, {Hathi},
  {Koekemoer}, {Kurczynski}, {Newman}, {Pacifici}, {P{\'e}rez-Gonz{\'a}lez}, \&
  {Pforr}}]{Salmon2016}
{Salmon}, B., {Papovich}, C., {Long}, J., {et~al.} 2016, \apj, 827, 20,
  \dodoi{10.3847/0004-637X/827/1/20}

\bibitem[{{Schreiber} {et~al.}(2018){Schreiber}, {Glazebrook}, {Nanayakkara},
  {Kacprzak}, {Labb{\'e}}, {Oesch}, {Yuan}, {Tran}, {Papovich}, {Spitler}, \&
  {Straatman}}]{Schreiber2018}
{Schreiber}, C., {Glazebrook}, K., {Nanayakkara}, T., {et~al.} 2018, \aap, 618,
  A85, \dodoi{10.1051/0004-6361/201833070}

\bibitem[{{Simha} {et~al.}(2014){Simha}, {Weinberg}, {Conroy}, {Dave},
  {Fardal}, {Katz}, \& {Oppenheimer}}]{Simha2014}
{Simha}, V., {Weinberg}, D.~H., {Conroy}, C., {et~al.} 2014, ArXiv e-prints.
\newblock \doarXiv{1404.0402}

\bibitem[{{Simpson} {et~al.}(2017){Simpson}, {Jimenez}, {Pena-Garay}, \&
  {Verde}}]{Simpson2017}
{Simpson}, F., {Jimenez}, R., {Pena-Garay}, C., \& {Verde}, L. 2017, \jcap, 6,
  029, \dodoi{10.1088/1475-7516/2017/06/029}

\bibitem[{Skilling(2006)}]{Skilling2006}
Skilling, J. 2006, Bayesian Anal., 1, 833, \dodoi{10.1214/06-BA127}

\bibitem[{{Speagle} {et~al.}(2014){Speagle}, {Steinhardt}, {Capak}, \&
  {Silverman}}]{Speagle2014}
{Speagle}, J.~S., {Steinhardt}, C.~L., {Capak}, P.~L., \& {Silverman}, J.~D.
  2014, \apjs, 214, 15, \dodoi{10.1088/0067-0049/214/2/15}

\bibitem[{{Taylor} {et~al.}(2011){Taylor}, {Hopkins}, {Baldry}, {Brown},
  {Driver}, {Kelvin}, {Hill}, {Robotham}, {Bland-Hawthorn}, {Jones}, {Sharp},
  {Thomas}, {Liske}, {Loveday}, {Norberg}, {Peacock}, {Bamford}, {Brough},
  {Colless}, {Cameron}, {Conselice}, {Croom}, {Frenk}, {Gunawardhana},
  {Kuijken}, {Nichol}, {Parkinson}, {Phillipps}, {Pimbblet}, {Popescu},
  {Prescott}, {Sutherland}, {Tuffs}, {van Kampen}, \&
  {Wijesinghe}}]{Taylor2011}
{Taylor}, E.~N., {Hopkins}, A.~M., {Baldry}, I.~K., {et~al.} 2011, \mnras, 418,
  1587, \dodoi{10.1111/j.1365-2966.2011.19536.x}

\bibitem[{{Thomas} {et~al.}(2017){Thomas}, {Le F{\`e}vre}, {Scodeggio},
  {Cassata}, {Garilli}, {Le Brun}, {Lemaux}, {Maccagni}, {Pforr}, {Tasca},
  {Zamorani}, {Bardelli}, {Hathi}, {Tresse}, {Zucca}, \&
  {Koekemoer}}]{Thomas2017}
{Thomas}, R., {Le F{\`e}vre}, O., {Scodeggio}, M., {et~al.} 2017, \aap, 602,
  A35, \dodoi{10.1051/0004-6361/201628141}

\bibitem[{{Tojeiro} {et~al.}(2007){Tojeiro}, {Heavens}, {Jimenez}, \&
  {Panter}}]{Tojeiro2007}
{Tojeiro}, R., {Heavens}, A.~F., {Jimenez}, R., \& {Panter}, B. 2007, \mnras,
  381, 1252, \dodoi{10.1111/j.1365-2966.2007.12323.x}

\bibitem[{{Tomczak} {et~al.}(2014){Tomczak}, {Quadri}, {Tran}, {Labb{\'e}},
  {Straatman}, {Papovich}, {Glazebrook}, {Allen}, {Brammer}, {Kacprzak},
  {Kawinwanichakij}, {Kelson}, {McCarthy}, {Mehrtens}, {Monson}, {Persson},
  {Spitler}, {Tilvi}, \& {van Dokkum}}]{Tomczak2014}
{Tomczak}, A.~R., {Quadri}, R.~F., {Tran}, K.-V.~H., {et~al.} 2014, \apj, 783,
  85, \dodoi{10.1088/0004-637X/783/2/85}

\bibitem[{{Trotta}(2008)}]{Trotta2008}
{Trotta}, R. 2008, Contemporary Physics, 49, 71,
  \dodoi{10.1080/00107510802066753}

\bibitem[{{Wild} {et~al.}(2016){Wild}, {Almaini}, {Dunlop}, {Simpson},
  {Rowlands}, {Bowler}, {Maltby}, \& {McLure}}]{Wild2016}
{Wild}, V., {Almaini}, O., {Dunlop}, J., {et~al.} 2016, \mnras, 463, 832,
  \dodoi{10.1093/mnras/stw1996}

\bibitem[{{Wright} {et~al.}(2016){Wright}, {Robotham}, {Bourne}, {Driver},
  {Dunne}, {Maddox}, {Alpaslan}, {Andrews}, {Bauer}, {Bland-Hawthorn},
  {Brough}, {Brown}, {Clarke}, {Cluver}, {Davies}, {Grootes}, {Holwerda},
  {Hopkins}, {Jarrett}, {Kafle}, {Lange}, {Liske}, {Loveday}, {Moffett},
  {Norberg}, {Popescu}, {Smith}, {Taylor}, {Tuffs}, {Wang}, \&
  {Wilkins}}]{Wright2016}
{Wright}, A.~H., {Robotham}, A.~S.~G., {Bourne}, N., {et~al.} 2016, \mnras,
  460, 765, \dodoi{10.1093/mnras/stw832}

\bibitem[{{Wright} {et~al.}(2017){Wright}, {Robotham}, {Driver}, {Alpaslan},
  {Andrews}, {Baldry}, {Bland-Hawthorn}, {Brough}, {Brown}, {Colless}, {da
  Cunha}, {Davies}, {Graham}, {Holwerda}, {Hopkins}, {Kafle}, {Kelvin},
  {Loveday}, {Maddox}, {Meyer}, {Moffett}, {Norberg}, {Phillipps}, {Rowlands},
  {Taylor}, {Wang}, \& {Wilkins}}]{Wright2017}
{Wright}, A.~H., {Robotham}, A.~S.~G., {Driver}, S.~P., {et~al.} 2017, \mnras,
  470, 283, \dodoi{10.1093/mnras/stx1149}

\bibitem[{{Wu} {et~al.}(2018){Wu}, {van der Wel}, {Gallazzi}, {Bezanson},
  {Pacifici}, {Straatman}, {Franx}, {Bari{\v s}i{\'c}}, {Bell}, {Brammer},
  {Calhau}, {Chauke}, {van Houdt}, {Maseda}, {Muzzin}, {Rix}, {Sobral},
  {Spilker}, {van de Sande}, {van Dokkum}, \& {Wild}}]{Wu2018}
{Wu}, P.-F., {van der Wel}, A., {Gallazzi}, A., {et~al.} 2018, \apj, 855, 85,
  \dodoi{10.3847/1538-4357/aab0a6}

\bibitem[{{Wuyts} {et~al.}(2009){Wuyts}, {Franx}, {Cox}, {Hernquist},
  {Hopkins}, {Robertson}, \& {van Dokkum}}]{Wuyts2009}
{Wuyts}, S., {Franx}, M., {Cox}, T.~J., {et~al.} 2009, \apj, 696, 348,
  \dodoi{10.1088/0004-637X/696/1/348}

\bibitem[{{Wuyts} {et~al.}(2011){Wuyts}, {F{\"o}rster Schreiber}, {Lutz},
  {Nordon}, {Berta}, {Altieri}, {Andreani}, {Aussel}, {Bongiovanni}, {Cepa},
  {Cimatti}, {Daddi}, {Elbaz}, {Genzel}, {Koekemoer}, {Magnelli}, {Maiolino},
  {McGrath}, {P{\'e}rez Garc{\'{\i}}a}, {Poglitsch}, {Popesso}, {Pozzi},
  {Sanchez-Portal}, {Sturm}, {Tacconi}, \& {Valtchanov}}]{Wuyts2011}
{Wuyts}, S., {F{\"o}rster Schreiber}, N.~M., {Lutz}, D., {et~al.} 2011, \apj,
  738, 106, \dodoi{10.1088/0004-637X/738/1/106}

\end{thebibliography}

\end{document}